\documentclass{article}
\usepackage[utf8]{inputenc}
\usepackage[T1]{fontenc}
\usepackage{fullpage}

\usepackage[round]{natbib}

\usepackage[margin = 1in]{geometry}

\usepackage{url}            %
\usepackage{booktabs}       %
\usepackage{amsfonts}       %
\usepackage{microtype}      %
\usepackage{wrapfig}

\usepackage{authblk}  
\usepackage{subcaption} 
\usepackage{times}
\usepackage{adjustbox}
\usepackage{dsfont}
\usepackage{amsmath, amssymb}
\usepackage{amsthm, thmtools, thm-restate}
\usepackage{bbm}
\usepackage{graphicx}
\usepackage{latexsym}
\usepackage{mathtools}
\usepackage{multirow}
\usepackage{paralist}
\usepackage{minitoc} %
\usepackage{xspace}
\usepackage{enumitem}
\usepackage{array}

\usepackage{tikz}
\usepackage{pgfplots}
\pgfplotsset{compat=1.18}
\usetikzlibrary{shapes}
\usetikzlibrary{positioning}
\usetikzlibrary{plotmarks}
\usetikzlibrary{patterns}
\usetikzlibrary{intersections,shapes.arrows}
\usetikzlibrary{pgfplots.fillbetween}

\usepackage[colorlinks,citecolor=bluegray,linkcolor=blue,urlcolor=blue,breaklinks]{hyperref}
\usepackage[capitalise]{cleveref}

\definecolor{orangeCB}{HTML}{FE6100}
\definecolor{maroonCB}{HTML}{DC267F}
\definecolor{blueCB}{HTML}{648FFF}
\definecolor{purpleCB}{HTML}{785ef0}
\definecolor{bluegray}{rgb}{0.04,0,0.7}
\definecolor{darkbrown}{rgb}{0.40,0.2,0.05}

\crefname{section}{Section}{Sections}

\crefname{theorem}{Theorem}{Theorems}
\crefname{remark}{Remark}{Remarks}
\crefname{lemma}{Lemma}{Lemmas}
\crefname{corollary}{Corollary}{Corollaries}
\crefname{proposition}{Proposition}{Propositions}
\crefname{assumption}{Assumption}{Assumptions}
\crefname{appendix}{Appendix}{Appendices}

\crefname{figure}{Figure}{Figures}
\crefname{table}{Table}{Tables}

\newtheorem{assumption}{Assumption}
\newtheorem{corollary}{Corollary}
\newtheorem{lemma}{Lemma}
\newtheorem{theorem}{Theorem}
\theoremstyle{definition}
\newtheorem{remark}{Remark}
\newtheorem{example}{Example}

\DeclareMathOperator*{\argmin}{arg\,min}
\DeclareMathOperator{\Tr}{Tr}
\newcommand{\Eb}[1]{\mathbb{E}\left[ #1 \right]}
\newcommand{\E}[1]{\mathbb{E}\left( #1 \right)}
\newcommand{\deriv}[2]{\frac{\partial{#1}}{\partial{#2}}}

\newcommand{\bX}{\mathbf{X}}
\newcommand{\bx}{\mathbf{x}}

\newcommand{\bz}{\mathbf{z}}
\newcommand{\bZ}{\mathbf{Z}}
\newcommand{\bzbar}{\overline{\mathbf{z}}}

\newcommand{\bY}{\mathbf{Y}}

\newcommand{\hGamma}{\hat{\Gamma}}
\newcommand{\tilGamma}{\tilde{\Gamma}}
\newcommand{\hgamma}{\hat{\gamma}}
\newcommand{\tilgamma}{\tilde{\gamma}}
\newcommand{\LAR}{L_{\mathrm{AR}}} 
\newcommand{\LVAR}{L_{\mathrm{VAR}}}
\newcommand{\hLVAR}{\hat{L}_{\mathrm{VAR}}} 

\newcommand{\VAR}{\mathrm{VAR}} 
\newcommand{\Var}{\mathrm{Var}} 
\newcommand{\Cov}{\mathrm{Cov}}  

\doparttoc
\faketableofcontents

\title{Order Selection in Vector Autoregression by Mean Square Information Criterion}
\author[1]{Michael Hellstern}
\author[1]{Ali Shojaie}

\affil[1]{Department of Biostatistics, University of Washington}
\date{}

\begin{document}
\maketitle

\begin{abstract}
Vector autoregressive (VAR) processes are ubiquitously used in economics, finance, and biology. Order selection is an essential step in fitting VAR models. While many order selection methods exist, all come with weaknesses. Order selection by minimizing AIC is a popular approach but is known to consistently overestimate the true order for processes of small dimension. On the other hand, methods based on BIC or the Hannan-Quinn (HQ) criteria are shown to require large sample sizes in order to accurately estimate the order for larger-dimensional processes. We propose the mean square information criterion (MIC) based on the observation that the expected squared error loss is flat once the fitted order reaches or exceeds the true order. MIC is shown to consistently estimate the order of the process under relatively mild conditions. Our simulation results show that MIC offers better performance relative to AIC, BIC, and HQ under misspecification. This advantage is corroborated when forecasting COVID-19 outcomes in New York City. Order selection by MIC is implemented in the \texttt{micvar} \texttt{R} package available on CRAN.
\end{abstract}

\section{Introduction} \label{sec:intro}
In vector autoregressive (VAR) models, each variable is modeled as a linear function of the multivariate time series over prior lags. VARs were first introduced in macroeconometrics by  \cite{sims1980macroeconomics} and have since become standard for macroeconomic forecasting. They have also become essential tools in a range of other fields, including in biomedical applications; in neuroscience to analyze functional connectivity in the brain \citep{seth2015granger} and in epidemiology to predict COVID-19 cases \citep{kitaoka2023improved}. A fundamental problem in fitting VAR models is how to choose the lag order: using too few lags may result in underfitting, while too many lags can lead to overfitting, both decreasing  the accuracy of forecasts. Incorrect selection of the lag order can also impact the selection of the relevant variables in the VAR model, resulting in ambiguous interpretations, especially when the goal is to infer Granger causal effects \citep{shojaie2022granger}. Unfortunately, the lag order is typically unknown and must be chosen either by prior knowledge or in a data-dependent way. 

\begin{figure}[t]
    \centering        
\includegraphics[width=0.9\linewidth]{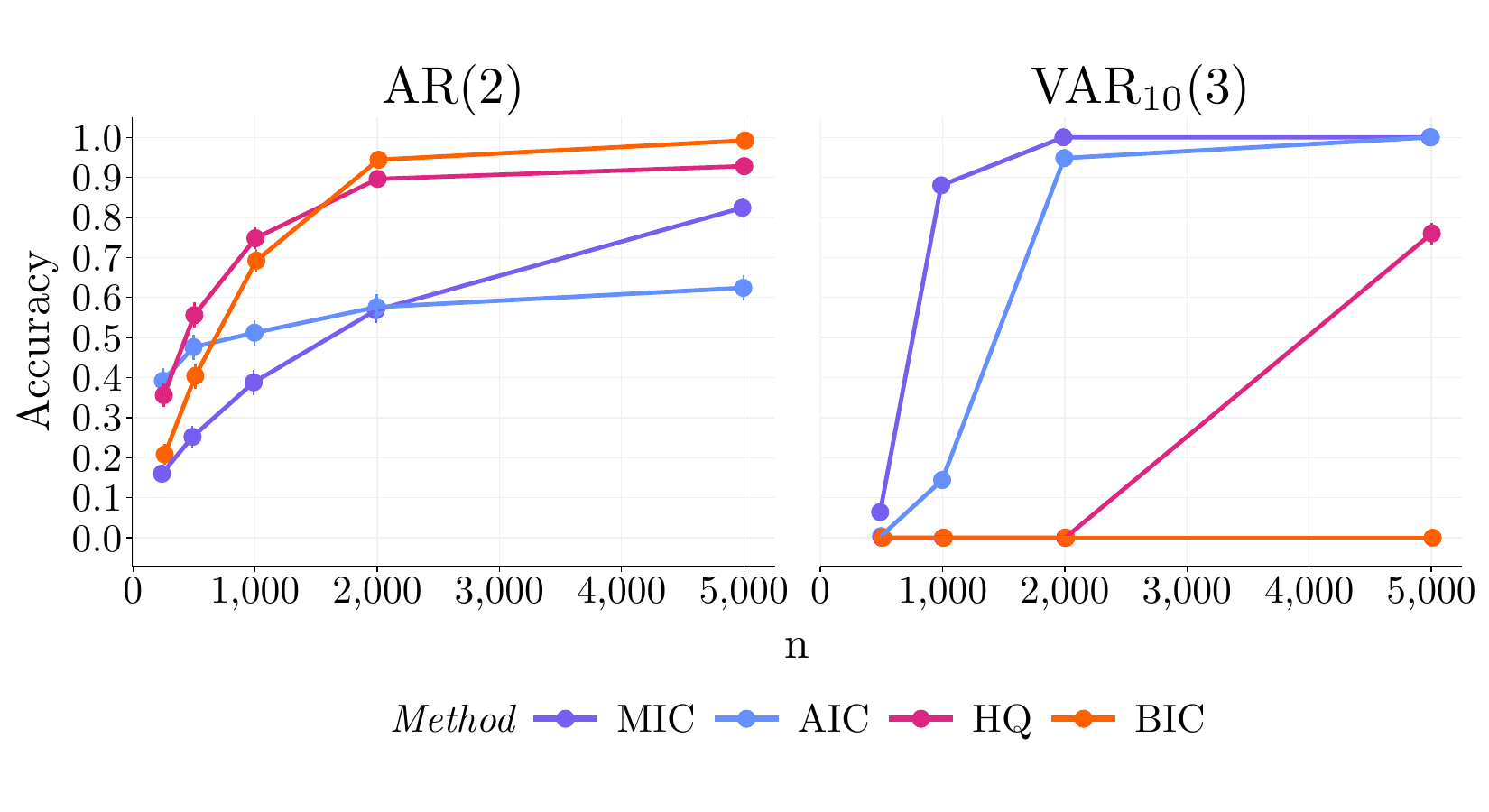}
    \caption[Accuracy of order selection methods]{Accuracy of various order selection methods in detecting the order of simulated $\mathrm{AR}(2)$ (\emph{left}) and $\mathrm{VAR}_{10}(3)$ (\emph{right}) processes. Vertical lines represent standard errors. Accuracy is measured as the proportion of simulations where the correct order was chosen. See \cref{sec:sims} for more details.}
    \label{fig:accuracy_var}
\end{figure}

Perhaps the most popular VAR order selection method is to choose the order that minimizes an information theoretic criterion, commonly the Akaike's Information Criterion (AIC) \citep{akaike1973information, akaike1974new}. Minimizing the AIC selects the model with the lowest negative log likelihood plus a penalty term on the number of independently adjustable parameters. Despite its popularity, AIC has several drawbacks for use in VAR models. The first stems from AIC's inherent reliance on the likelihood. In the context of VAR models, this often amounts to assuming a Gaussian likelihood, which results in simplifying the log likelihood term to the log determinant of the prediction error matrix. When the errors are not Gaussian, this simplification of the likelihood is no longer valid. 
The second is that AIC may not provide a consistent estimate of the VAR order \citep[ Corollary 4.2.1]{lutkepohl2005new}. Although this limitation improves as the dimension of the process increases and is negligible for VAR models of dimension greater than 5 \citep{paulsen1985estimation}, it nonetheless limits the applicability of AIC.  
This lack of consistency is highlighted in the simulation results in the left panel of \cref{fig:accuracy_var}. The plot shows that in the univariate case, AIC reaches its peak accuracy of 0.6 at around $n=2000$ and does not improve substantially as $n$  increases to $5000$. These results are in contrast to those presented in the right panel of \cref{fig:accuracy_var} for a 10 dimensional VAR model: in this case, AIC has nearly perfect accuracy for $n \geq 2000$. 
Shortly after AIC was introduced, additional VAR order selection criteria were proposed, namely the Bayesian information criterion, \citep[BIC, ][]{schwarz1978estimating} and the Hannan-Quinn (HQ) criterion \citep{hannan1979determination, quinn1980order}. Similar to AIC, the BIC criterion relies on the likelihood, which in the case of Gaussian errors amounts to the log determinant of the prediction error, plus a penalty on the number of model parameters. The HQ criterion is not likelihood-based but it too relies on the log determinant of the prediction error plus a penalty. Although all three criteria use the log determinant of the prediction error, they differ in the penalty used. Denoting by $n$ the number of observations, AIC uses a penalty of $(2/n)(\mathrm{\# parameters})$, while HQ and BIC use penalties of $((2 \log \log n)/n)(\mathrm{\# parameters})$ and $((\log n)/n)(\mathrm{\# parameters})$, respectively. This change of penalty is essential: if the underlying process is a stationary and stable VAR with standard white noise, it can be shown that both HQ and BIC consistently estimate the true order of the process \citep[Corollary 4.2.2]{lutkepohl2005new}. However, while HQ and BIC are consistent, simulation results in \cref{fig:accuracy_var} show they perform poorly for small sample sizes when the dimension of the process increases to a moderate size.

To address the above limitations of existing order selection methods, we propose a new method, mean squared information criterion (MIC). MIC is likelihood-free, consistent, and performs well in a variety of simulation settings. Our criterion leverages the novel observation that the expected squared error loss is constant when the fitted VAR order is at least as large as the true model order. We establish the consistency of MIC under mild assumptions and show that, compared with AIC, HQ, and BIC, it performs well in a variety of simulation settings and forecasting on real data.

The rest of the paper is structured as follows. In \cref{sec:new_idea}, we motivate our new information criterion and show that the expected squared error loss is constant when the fitted order is at least as large as the true model order. We extend this observation to the multivariate case and present theoretical results in \cref{sec:multivariate_ext}, introduce our estimator in \cref{sec:order_estimator} and compare its performance to AIC, HQ and BIC using simulated data in \cref{sec:sims}. We apply our method to a financial application and COVID-19 forecasting in \cref{sec:applications} and end with a discussion in \cref{sec:discussion}.

\textit{Notation:} We use uppercase letters $X, Y, Z$ to denote random variables. We will use subscript $t$ to denote the time component as in $Z_t$. When $Z$ or $Z_t$ are random vectors, the components can be accessed by $Z_j$, $Z_{t,j}$, respectively. Observed values of random variables are written in lowercase as in $z_t$. Additionally, observed vectors and matrices are denoted using bold lowercase and uppercase letters as in $\bz$ and $\bZ$, respectively. Hats or tildes will be used to specify sample estimates of population quantities, e.g. $\hGamma_0$ and $\tilGamma_0$ represent different sample estimates of $\Gamma_0$.

\section{A new idea}\label{sec:new_idea}

We begin with the univariate case. Let $Z_t$ be a stationary univariate time series. Without loss of generality, we assume  $Z_t$ is a mean zero process, as in practice, we can subtract the sample mean from the data. We denote the $h^{\text{th}}$ autocovariance of $Z_t$ as $\gamma_h = \E{Z_{t}Z_{t-h}}$ and denote $Y_t = Z_t, X_{p} = \begin{bmatrix} Z_{t-1} \dots Z_{t-p} \end{bmatrix}$. We wish to study the behavior of the expected squared error loss as a function of different orders $p$ up to a prespecified maximum order, $p_{\max}$. Specifically, we study
\begin{equation*}
    \LAR(p, \beta) := \Eb{\left(Y_t - X_{p} \beta \right)^2} \, .
\end{equation*}
In this case, $\beta$ is a nuisance parameter. For fixed $p$, $\LAR(p, \beta)$ is the usual least squares problem and we can solve for $\beta$ to get $\beta^*_p = \E{X_p^T X_p}^{-1} \E{X_p^T Y_t}$. Plugging $\beta^*_p$ back in to the expected loss and simplifying, we get a loss that only depends on $p$, for which we use the shorthand notation $\LAR(p)$:
\begin{equation*}
    \LAR(p) = \E{Y_t^2} - \E{X_p^T Y_t}^T \E{X_p^T X_p}^{-1} \E{X_p^T Y_t} \, .
\end{equation*}
By stationarity of $Z_t$, this simplifies to
\begin{equation*}
    \LAR(p) =  \gamma_0 - \begin{bmatrix} \gamma_1 & \dots & \gamma_p \end{bmatrix}  \begin{bmatrix}
        \gamma_0 & \gamma_1 & \gamma_2 & \dots \\
        \gamma_1 & \gamma_0 & \gamma_1 & \dots \\
        \gamma_2 & \gamma_1 & \gamma_0 & \dots \\
        \vdots & \vdots & \vdots & \ddots & \\
        \gamma_{p-1} & \gamma_{p-2} & \gamma_{p-3} & \dots & \gamma_0
    \end{bmatrix}^{-1}
    \begin{bmatrix} \gamma_1 \\ \vdots \\ \gamma_p \end{bmatrix} \, .
\end{equation*}
Note that up to this point we have only assumed $Z_t$ is stationary and have made no assumptions on the structure of the data generating process. Now suppose $Z_t$ is not only stationary but is also a stable $\mathrm{AR}(p_0)$ process, that is,
\begin{equation*}
    Z_t = a_1 Z_{t-1} + \dots + a_{p_0}Z_{t-p_0} + \epsilon_t, 
\end{equation*}
where $\E{\epsilon_t} = 0$, $\E{\epsilon_t^2} = \sigma^2_{\epsilon}$ and $\E{\epsilon_s \epsilon_t} = 0$ for $s \neq t$.  Then, $\gamma_h$ has a known form and we can calculate the expected squared error loss for each $p = 0, \dots, p_{\max}$ \citep[pp. 26-30]{lutkepohl2005new}. For $p = 0$, there are no prior lags as predictors so we get that $\LAR(0) = \gamma_0 $.

\begin{figure}[t]
    \includegraphics[width=0.5\textwidth]{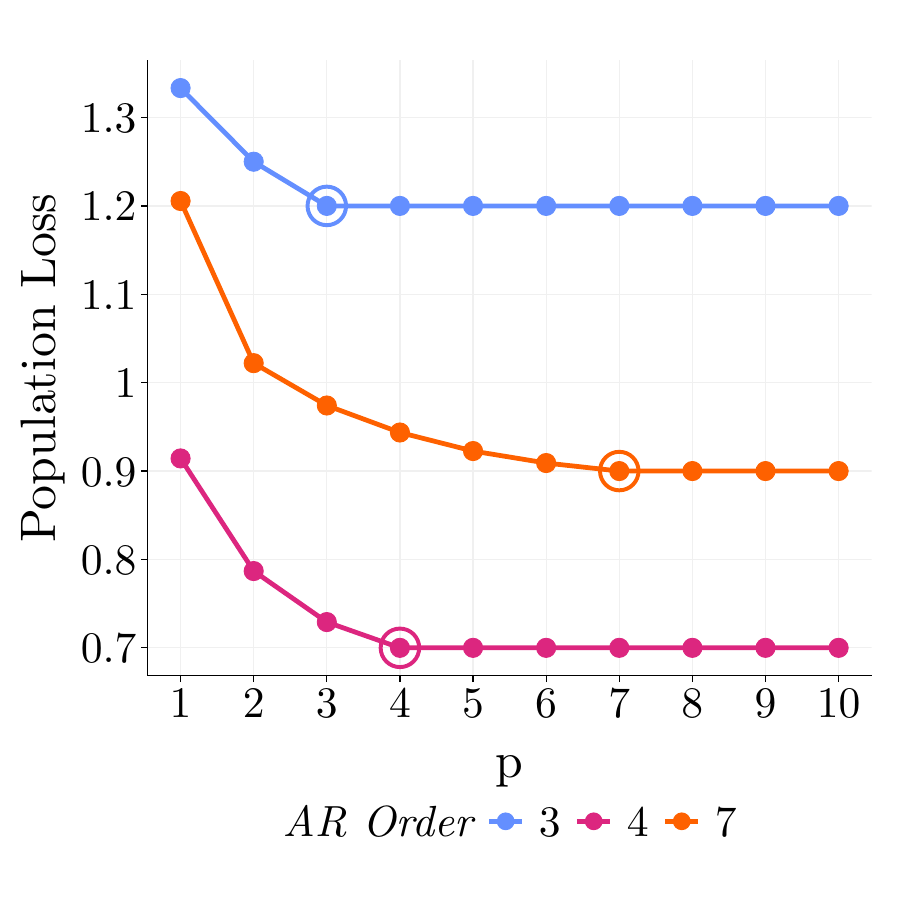}
    \centering
    \caption[Population loss for AR processes]{Population loss for AR(3), AR(4), and AR(7) processes with $\sigma_{\epsilon}^2 = 1.2, 0.7, 0.9$ respectively. As the fitted order increases to the true order the expected loss decreases monotonically to $\sigma^2_{\epsilon}$. The true order for each line is denoted by a hollow circle.}
    \label{fig:pop_loss}
\end{figure}

We now observe that the expected squared error loss should be flat after the true order $p_0$. This is because the first $p_0$ lags should contain all the information needed for prediction. Furthermore, the error of the process is white noise and thus unpredictable. Therefore, the lowest achievable expected squared error should be the variance of the error. In \cref{fig:pop_loss}, we compute the expected squared error loss for several AR processes and see that our intuition holds. It can be seen that the expected squared error loss is indeed flat once the fitted order is at least as large as the true order $p_0$ and the eventual value of the expected loss is the variance of the error, $\sigma^2_{\epsilon}$.  
The behavior of the loss observed in \cref{fig:pop_loss} can be proven mathematically and follows immediately from the multivariate case proved in \cref{thm:flat_loss} in the next section. Given this behavior, if we add an appropriately sized penalty to the fitted order $p$ and consider orders large enough ($p_{\max} > p_0$), we should be able to design a correct order selection procedure in the sense that when the process is AR($p_0$) we will recover the true order $p_0$. Specifically, the true order can be found as 
\begin{equation*}
    p_0 = \argmin_{p \in \{0, \dots, p_{\max}\}} \LAR(p) + \lambda p \, .
\end{equation*}
If $\lambda$ is too large, we would  underestimate the order. However, we must also have $\lambda > 0$ to avoid multiple solutions and an undefined parameter. In practice, we rarely deal with univariate time series so in the next section we extend these concepts to the multivariate case.

\section{Extension to the multivariate case}\label{sec:multivariate_ext}

In this section, we extend the concepts of \cref{sec:new_idea} to the multivariate case and show a similar behavior of the expected squared error loss. Suppose $Z_t = \begin{bmatrix} Z_{t,1}, \dots, Z_{t,k}\end{bmatrix}^T$ is a $k$-dimensional column vector. We assume that $Z_t$ is a stable process, as formalized in \cref{asm:stability} (stability) in \cref{app:assumptions}. The $h^{\text{th}}$ autocovariance matrix is defined as $\Gamma_{h} = \E{Z_{t} Z_{t-h}^T} \in \mathbb{R}^{k \times k}$. Note that $\Gamma_h$ is not symmetric in general, but $\Gamma_h = \Gamma_{-h}^T$. Similar to the univariate case, we define $Y_t = Z_t, X_p = \begin{bmatrix} Z_{t-1}^T \dots Z_{t-p}^T \end{bmatrix}^T$, where $Y_t \in \mathbb{R}^{k \times 1}$ and $X_p \in \mathbb{R}^{kp \times 1}$. We study the expected squared error loss at different values of $p$,
\begin{equation*}
    \LVAR(p, A_p) := \Eb{ \left(Y_t - A_pX_p \right)^T\left(Y_t - A_pX_p \right) } \, .
\end{equation*}
We proceed by profiling out $A_p$. For fixed $p$, we have $A^*_p = \E{Y_t X_{p}^T }\E{X_{p} X_{p}^T}^{-1}$. Similar to the univariate case, we plug $A^*_p$ back into the expected loss which we denote as $\LVAR(p)$ after simplification. It is shown in \cref{app:simplify_loss} that
\begin{equation}
\label{eqn:LVAR_autocov}
    \LVAR(p) = \Tr\left( \Gamma_0 \right) - 
    \Tr \left( 
    \begin{bmatrix} \Gamma_1 & \dots & \Gamma_p \end{bmatrix}  
    \begin{bmatrix} \Gamma_0 & \dots & \Gamma_{p-1} \\ 
                                    \vdots & & \vdots \\
                                    \Gamma_{p-1}^T & \dots & \Gamma_0 
    \end{bmatrix}^{-1}  
    \begin{bmatrix} \Gamma_1^T \\ \vdots \\ \Gamma_p^T \end{bmatrix}
    \right) \, .
\end{equation}
For $p = 0$ there are no prior lags as predictors so we get that $\LVAR(0) = \Tr\left( \Gamma_0 \right) $. Similar to the univariate case, if $Z_t$ is a stable VAR$(p_0)$ process, the multivariate loss decreases until the fitted order is equal to the true order at which point the population loss is constant with a value of $\Tr(\Sigma_{\epsilon})$. This behavior of the loss is formally stated in \cref{thm:flat_loss}. The assumptions are stated in Appendix~\ref{app:assumptions}

\begin{theorem}[Flat Loss] \label{thm:flat_loss}
    Suppose $Z_t$ is a $\VAR(p_0)$ process with standard white noise. That is, $Z_t = \sum_{i=1}^{p_0} A_i Z_{t-i} + \epsilon_t$ where $\E{\epsilon_t} = 0, \E{\epsilon_t \epsilon_t^T} = \Sigma_{\epsilon}$ and $\E{\epsilon_t \epsilon_s^T} = 0$ for $t \neq s$. If \cref{asm:stability} (stability), \cref{asm:invertibility} (invertibility), and \cref{asm:irreducibility} (irreducibility) hold, then 
    \begin{equation*}
        \begin{cases}
        \LVAR(p) < \LVAR(p-1) & \text{ if } p \leq p_0 \\
        \LVAR(p) = \Tr(\Sigma_{\epsilon}) & \text{ if } p \geq p_0 \, .
        \end{cases}
    \end{equation*}
\end{theorem}

\cref{asm:stability,asm:invertibility,asm:irreducibility} are mild and standard assumptions which hold in many applications. Further discussion on the assumptions is provided in \cref{app:assumptions}. For $\LVAR(p_0)$, \cref{thm:flat_loss} implies both $\LVAR(p_0) < \LVAR(p_0 - 1)$ and $\LVAR(p_0) = \Tr(\Sigma_{\epsilon})$. If we penalize the fitted order $p$ by an appropriate amount, and use $p_{\max} > p_0$, we would obtain a procedure that recovers the true order, $p_0$. This is formally stated in \cref{cor:param_defn}.
\begin{corollary}\label{cor:param_defn}
    Let $M = \min\left( \LVAR(p_0 - 1) - \LVAR(p_0), \left[\LVAR(p_0 - 2) - \LVAR(p_0 ) \right]/2, \dots, \left[ \LVAR(0) - \LVAR(p_0)\right]/p_0 \right)$. Then, for a VAR$(p_0)$ process that satisfies the conditions of \cref{thm:flat_loss} and for $\lambda \in (0, M)$ and $p_{\max} > p_0$ we have
    \begin{equation*}
    p_0 = \argmin_{p \in \{0, \dots, p_{\max}\}} \LVAR(p) + \lambda p \, .
    \end{equation*}
\end{corollary}

To estimate $p_0$, we need to estimate $\LVAR(p) + \lambda p$ for each $p$.While we use the form $\LVAR(p)$ with the autocovariance matrices to establish the theoretical results, $\LVAR(p)$ can be expressed as the expected squared error loss:  $\LVAR(p) = \mathbb{E}\left[ \left(Y_t - A^*_p X_p \right)^T\left(Y_t - A^*_p X_p \right) \right]$. Therefore, a natural estimator of $\LVAR(p)$ is the sample squared error loss, as defined formally below.

Let $\{ \bz_t \}_{t = 1}^{n}$ denote the observed $k$-dimensional time series of length $n$. To estimate the squared error loss based on the least squares estimate of $A^*_p$ at a fitted order $p$, we denote $\bY_p = \begin{bmatrix} \bz_{1+p}, \dots, \bz_n \end{bmatrix} \in \mathbb{R}^{k \times (n - p)}$, $\bx_{t,p} = \begin{bmatrix} \bz_{t - 1}^T & \dots & \bz_{t - p}^T \end{bmatrix}^T \in \mathbb{R}^{kp \times 1}$ and $\bX_{p} = \begin{bmatrix} \bx_{1+p,p} & \dots & \bx_{n,p} \end{bmatrix} \in \mathbb{R}^{kp \times (n-p)}$. The number of data points in $\bX_p$ and $\bY_p$ depends on the fitted order $p$ as fitting a $\VAR(p)$ model requires the $p$ prior data points as covariates. As a result, we only have $n - p$ usable data points for a $\VAR(p)$ model. With these definitions, an estimate of the squared error loss at a fitted order $p$ using the least squares estimates of $A^*_p$ is given by
\begin{equation}\label{eqn:sampleloss}
    \begin{split}
        \hLVAR(p) & =  \frac{1}{n-p}  \Tr\left(  \left( \bY_p - \hat{A}_p \bX_p \right)^T\left( \bY_p - \hat{A}_p \bX_p \right) \right) \\
        & = \frac{1}{n-p}  \Tr\left(  \left( \bY_p - \hat{A}_p \bX_p \right)\left(\bY_p - \hat{A}_p \bX_p \right)^T \right), 
    \end{split}
\end{equation}
where $\hat{A}_p = \bY_p \bX_p^T \left( \bX_p \bX_p^T \right)^{-1}$ and we used $\Tr(A^TB) = \Tr(AB^T)$. 

With \eqref{eqn:sampleloss} and using $\lambda \in (0, M)$, we can  define the estimated order based on the mean-squared information criterion (MIC), $p^*_{\mathrm{MIC}}$, as
\begin{equation} \label{eqn:MIC_lamTrue}
p^*_{\mathrm{MIC}} = \argmin_{p \in \{0, \dots, p_{\max}\}} \hLVAR(p) + \lambda p \,.
\end{equation} 
Note that to use this estimator in practice, we need to specify the tuning parameter $\lambda$. We defer this to \cref{eqn:MIC_est} below. To solve the minimization problem in \cref{eqn:MIC_lamTrue}, we can compute $\hLVAR(p)$ for each $p = 0, \dots, p_{\max}$. \cref{thm:consistency} establishes the consistency of $\hLVAR(p)$.
\begin{theorem}[Consistency of Loss] \label{thm:consistency}
    Under the assumptions of \cref{thm:flat_loss}, we have that 
\[
    \left| \hLVAR(p) - \LVAR(p) \right| = o_P(n^{-1/2 + \delta}) \, ,
\]
    for all $\delta > 0$.
\end{theorem}
As shown in the proof of \cref{thm:consistency} (see \cref{app:consistent_sample_loss}), the consistency of $\hLVAR(p)$ relies on the consistency of the sample autocovariances and the rate of convergence of $\hLVAR(p)$ is the same as the rate of convergence of the sample autocovariance. One benefit of the proposed information criterion is that it does not rely on the form of the  likelihood. In fact, Theorem~\ref{thm:consistency} will hold as long as the sample autocovariance is consistent, with the rate of convergence matching that of the sample autocovariance. An immediate consequence of \cref{thm:consistency} is consistency of $p^*_{\mathrm{MIC}}$.

\begin{corollary}[Consistency of order estimate] \label{cor:consistency_order}
    Under the assumptions of \cref{thm:flat_loss}, we have that 
    \[
     p^*_{\mathrm{MIC}} \rightarrow_p p_0, 
    \]
    where $\rightarrow_p$ denotes convergence in probability.
\end{corollary}

While our theoretical analyses and implementation estimate $\LVAR(p)$ using the least squares estimate of $A^*_p$, it is possible to use other estimators, such as the Yule-Walker estimate of $A^*_p$, and prove similar results. However, we chose to use the least squares estimate due to better small sample performance and lower bias \citep{lutkepohl2005new, tjostheim1983bias}. It is worth noting that, as pointed out in \citet[pp. 86]{lutkepohl2005new}, the least squares and Yule-Walker estimators of $A^*_p$ are asymptotically equivalent for stable processes, which are the focus of this work.

\section{MIC estimator of VAR order} \label{sec:order_estimator}
Our theoretical analyses in Section~\ref{sec:multivariate_ext} assume $\lambda \in (0, M)$ is known. However, $\lambda$ is unknown in practice and needs to be selected. The flat loss concept from \cref{thm:flat_loss} tells us that once the fitted order exceeds the true order, the loss should be constant. In practice, there is sampling variability so the loss will never completely stabilize. We generate an estimate of this variability using a ``self-tuning'' approach. 

In our self-tuning approach, we fit models from lag order $p_{\max}+1$ to $2p_{\max}$ and take the absolute value of the mean of the difference between each loss and the subsequent loss to estimate the amount of variability, or noise, in the loss. While it is possible to use orders larger than $2p_{\max}$, the trade-off is fitting larger order models consumes more prior data points as covariates and reduces sample size. We find that $2p_{\max}$ works well in practice.
Specifically, our self-tuning approach computes
\begin{equation*}
    \mathrm{MD} = \left| \mathrm{mean} \left( \hLVAR(p_{\max}) - \hLVAR(p_{\max}+1) , \dots, \hLVAR(2p_{\max}-1) - \hLVAR(2p_{\max}) \right) \right| \, .
\end{equation*} 
We then scale $\mathrm{MD}$ by $\sqrt{n / (k^2 \log(n))}$ to get 
\begin{equation*}
    \lambda_{\mathrm{ST}} = \mathrm{MD} \sqrt{\frac{n}{k^2 \log(n)}} \, .
\end{equation*}
With this choice of $\lambda$, our estimator is defined as
\begin{equation}
    \label{eqn:MIC_est}
    \hat{p}_{\mathrm{MIC}} = \argmin_{p \in \{0, \dots, p_{\max}\}} \hLVAR(p) + \lambda_{\mathrm{ST}} p \,.
\end{equation} 
We next discuss the choice of $\mathrm{MD}$ as well as the scaling $\sqrt{n / (k^2 \log(n))}$.

Due to the flat loss property, each $\hLVAR(p_{\max}+i) - \hLVAR(p_{\max}+i+1)$ should represent an estimate of how the sample loss changes when we have exceeded the true order and increase the fitted order by 1. We average over $p_{\max}$ of these to reduce the variance in this estimate. When computing the mean, subsequent differences cancel and this quantity can be simplified as
\begin{equation*}
    \mathrm{MD} = \left| \frac{\hLVAR(p_{\max}) - \hLVAR(2p_{\max}) }{p_{\max}}\right| \, .
\end{equation*}
Thus, $\mathrm{MD}$ can be computed efficiently as it only requires fitting one additional regression of order $2p_{\max}$. It is also worth noting that $\hLVAR(p_{\max}), \dots, \hLVAR(2p_{\max})$ converge to the same asymptotic distribution and are asymptotically perfectly correlated. In this instance, the correlation is beneficial as it further reduces the variance of our estimate since for two random variables, $X, Y$, $\Var(X - Y) = \Var(X) + \Var(Y) - 2 \Cov(X,Y)$.

To understand why it is necessary to scale $\mathrm{MD}$, consider the case where we instead use $\lambda_{\mathrm{ST}} = \mathrm{MD}$. To simplify notation and provide a more concrete setting, consider $p_{\max} = 10$. With these, the score for order $2p_{\max} := 20$ based on \cref{eqn:MIC_est} becomes
\begin{align}
\label{eqn:md_scale_ex}
    \hLVAR(20) + \frac{\hLVAR(10) - \hLVAR(20) }{10} 20  & = \frac{20}{10} \hLVAR(10)-\frac{10}{10}\hLVAR(20) \nonumber \\
    & = \hLVAR(10) + \frac{\hLVAR(10) - \hLVAR(20) }{10} 10 \, ,
\end{align}
where we have assumed that $\hLVAR(10) > \hLVAR(20)$ so we can ignore the absolute value in $\mathrm{MD}$. While it is possible that $\hLVAR(10) < \hLVAR(20)$ due to the different datasets used in fitting each model---e.g. $\hLVAR(10)$ is estimated using $n - 10$ observations while $\hLVAR(20)$ uses $n - 20$ observations---this is unlikely. 

The last equation on the right hand side of \cref{eqn:md_scale_ex} is exactly the value of the penalized loss for order $p_{\max} = 10$. In other words, setting $\lambda_{\mathrm{ST}} = \mathrm{MD}$ treats models $p_{\max}$ and $2 p_{\max}$ as equally viable. However, these models are not equally viable and we want to enforce a belief that higher orders are less likely. Thus, we need to choose a penalty that is larger than $\mathrm{MD}$. One way to do this is to scale $\mathrm{MD}$ by a factor greater than $1$. From \cref{thm:consistency}, we know that $\hLVAR(p) = \LVAR(p) + o_P(n^{-1/2 + \delta}) \quad \forall\ \delta > 0$ so scaling by $n^{1/2}$ is too fast. In practice, we find that $\sqrt{n/\log(n)}$ works well. Lastly, we scale by $1/k$ since each additional order fitted requires estimating $k$ more parameters than the prior order (see the proof of \cref{thm:flat_loss}).

\subsection{Relationship between MIC and other criteria}

We now compare our method, MIC, to AIC, BIC, and HQ. For ease of comparison, we write the estimated error covariance matrix as $\hat{\Sigma}_p$.
\begin{equation*}
    \hat{\Sigma}_p = \frac{1}{n-p}   \left( \bY_p - \hat{A}_p \bX_p \right)\left(\bY_p - \hat{A}_p \bX_p \right)^T. 
\end{equation*}
Note that $\Tr(\hat{\Sigma}_p)  = \hLVAR(p)$. Denoting the natural logarithm as $\log$, the criteria considered can be written as follows
\begin{align*}
    \mathrm{MIC}(p) & = \Tr(\hat{\Sigma}_p) + \mathrm{MD} \sqrt{\frac{n}{k^2 \log(n)}} p & \qquad 
     \mathrm{AIC}(p) & = \log \left|\hat{\Sigma}_p \right| + \frac{2}{n} k^2p \, ; \\
     \mathrm{BIC}(p) & = \log \left|\hat{\Sigma}_p \right| + \frac{ \log n}{n} k^2 p & \qquad \mathrm{HQ}(p) &= \log \left|\hat{\Sigma}_p \right| + \frac{2  \log \log n}{n} k^2 p \, .
\end{align*} 
The above formulations show that all criteria rely on the same estimate of the error matrix $\hat{\Sigma}_p$ and differ only in how they use it---$\Tr()$ or $\log | \cdot |$---and penalize that information. The advantage of MIC is that it is consistent and likelihood-free. Our simulation results also show that MIC works well empirically.

\subsection{Alternative choices of \texorpdfstring{$\lambda$}{lambda}}
\label{sec:alt_lambdas}

We also considered two other methods to select the order based on the flat loss concept discussed in Section~\ref{sec:multivariate_ext}. The first method, which we refer to as MIC-sp, uses MIC with a penalty $\lambda_{\mathrm{sp}}$ that is chosen by splitting the data into train and test sets. Due to the time-dependent nature of our data, we use the first 70\% of observations as training and the remainder as test data. Models from $p_{\mathrm{max}}$ to $2p_{\mathrm{max}}$ are fit on the training data to estimate $\hat{A}_{p}$ and their prediction errors computed on the test data are denoted as e.g. $\hat{\Sigma}_{\mathrm{test}, p_{\max}}$. We set 
\begin{equation*}
    \lambda_{\mathrm{sp}} = \mathrm{mean} \left( \left|\Tr(\hat{\Sigma}_{\mathrm{test}, p_{\max}}) - \Tr(\hat{\Sigma}_{\mathrm{test}, p_{\max} + 1}) \right|, \dots, \left|\Tr(\hat{\Sigma}_{\mathrm{test}, 2p_{\max} - 1}) - \Tr(\hat{\Sigma}_{\mathrm{test}, 2p_{\max}}) \right| \right) \, .
\end{equation*}
Due to the flat loss property, each of these differences should be 0 and any sample variability should be captured in $\lambda_{\mathrm{sp}}$. 

We further consider a procedure, which we denote as MIC-mt. We again use a 70-30 train-test split and fit VAR models of order $0, \dots, p_{\mathrm{max}}$ on the train dataset. Similarly, we compute the errors of each fitted model on the test data. MIC-mt then chooses the order $0, \dots, p_{\mathrm{max}}$ that minimizes the test error. Simulations comparing all three methods in the case of a diagonal covariance matrix with Gaussian errors are shown in \cref{fig:MICcomp_diagGaus_sims}. The results show that MIC, which indicates the MIC method with self-tuned $\lambda$, performs the best across a variety of sample sizes and dimensions. It is only consistently outperformed by MIC-sp in the AR(2) case. Results for other simulations in \cref{sec:sims} are not shown but are qualitatively similar. Thus, we proceed with our proposed self-tuning approach for selecting $\lambda$.

\section{Simulations}\label{sec:sims}
\subsection{Order selection accuracy}

In this section, we compare the accuracy of MIC, AIC, BIC, and HQ order selection methods using simulated data. In general, we will use $\VAR_{k}(p)$ to denote the dimension $k$ and order $p$ of the process. We will also use U$(a,b)$ to denote a Uniform distribution with support $(a,b)$. We consider VAR models with 4 different dimensions and 3 different error structures. The first is an autoregressive process of order 2, AR(2), with parameters (0.3, 0.1). The second is a $\VAR_{2}(2)$ process. The entries of the first lag coefficient matrix are 25\% sparse and randomly drawn from either a U$(0.1,0.3)$ or a U$(-0.3,-0.1)$ each with 50\% probability. The entries of the second lag coefficient matrix are 50\% sparse and randomly drawn from either a U$(0.07,0.2)$ or a U$(-0.2,-0.07)$ each with 50\% probability. The third simulation setting is a $\VAR_{5}(3)$. All lag coefficient matrices have 60\% sparsity. In the first lag coefficient matrix, the non-zero entries are drawn from a U$(0.1,0.3)$ or a U$(-0.3,-0.1)$ each with equal probability. The second lag coefficient matrix uses a U$(0.1,0.2)$ or a U$(-0.2,-0.1)$ while the third uses a U$(0.05,0.1)$ or a U$(-0.1,-0.05)$. The fourth simulation setting is a $\VAR_{10}(3)$ process where the first lag coefficient matrix has 40\% sparsity and the non-zero entries are drawn from a U$(0.1,0.3)$ or a U$(-0.3,-0.1)$. The second lag coefficient matrix has 80\% sparsity, but the remaining entries are drawn from a U$(-0.2,0.2)$. The final lag coefficient matrix has 80\% sparsity with remaining entries drawn from a U$(-0.1,0.1)$. All coefficient matrices are generated once and the same matrices are used throughout the simulations. Stability for each setting is verified using the method of \citet[pp. 14 - 17]{lutkepohl2005new}.

All datasets are simulated using three error structures. The first is mean-zero Gaussian errors with an identity covariance matrix while the second uses a randomly generated covariance matrix. The covariance matrix is generated by generating a $k \times k$ matrix with entries drawn from a U$(-3,3)$. The matrix is then symmetrized by left multiplying by its transpose. We enforce a maximum condition number of 100 for each matrix by consecutively adding $0.001$ to diagonal elements until the condition number is met. After the covariance matrix is reconditioned, it is scaled to have unit variances. The third error structure is a Gaussian mixture model with $5$ components. Each component is Gaussian where the mean vector is generated from a U$(-5,5)$. For each $k$, we then subtract the mean across all $5$ components so that the mean of the component means is 0. The covariances are $k \times k$ matrices with entries drawn from a U$(-3,3)$. The matrices are subsequently symmetrized, reconditioned, and rescaled as explained above. We simulate $n = 250, 500, 1000, 2000, 5000$ observations for each setting except for $\VAR_{10}(3)$ where $n = 250$ is excluded as the number of parameters exceeds the number of data points.  

Lastly, we consider a $\VAR_{3}(2)$ process that switches between one of two regimes. Both regimes share the same lag coefficient matrices and error covariances but differ in their means. Since both regimes are order 2, the true order is 2. The regime mean vectors are generated from U$(-0.5,0.5)$ and the regime switches every 10\% of observations. For example, if $n = 5000$, there are nine regime swtiches at $n = 500, 1000, 1500, \dots, 4500$. Due to the time-dependent switching mean, this process is not stationary and we present these results to study how the methods perform under misspecification. For this process, the first lag coefficient matrix is 30\% sparse with entries randomly drawn from a U$(0.1,0.3)$ or U$(-0.3,-0.1)$ while the second is 60\% sparse with entries drawn from a U$(0.1,0.2)$ or U$(-0.2,-0.1)$. The errors are generated from a Gaussian distribution with mean zero and randomly generated covariance matrix as above. Prior to analyzing the data, each of the three process components is centered so the overall mean of the components is 0.

\begin{figure}
\centering
\includegraphics[width=0.75\linewidth]{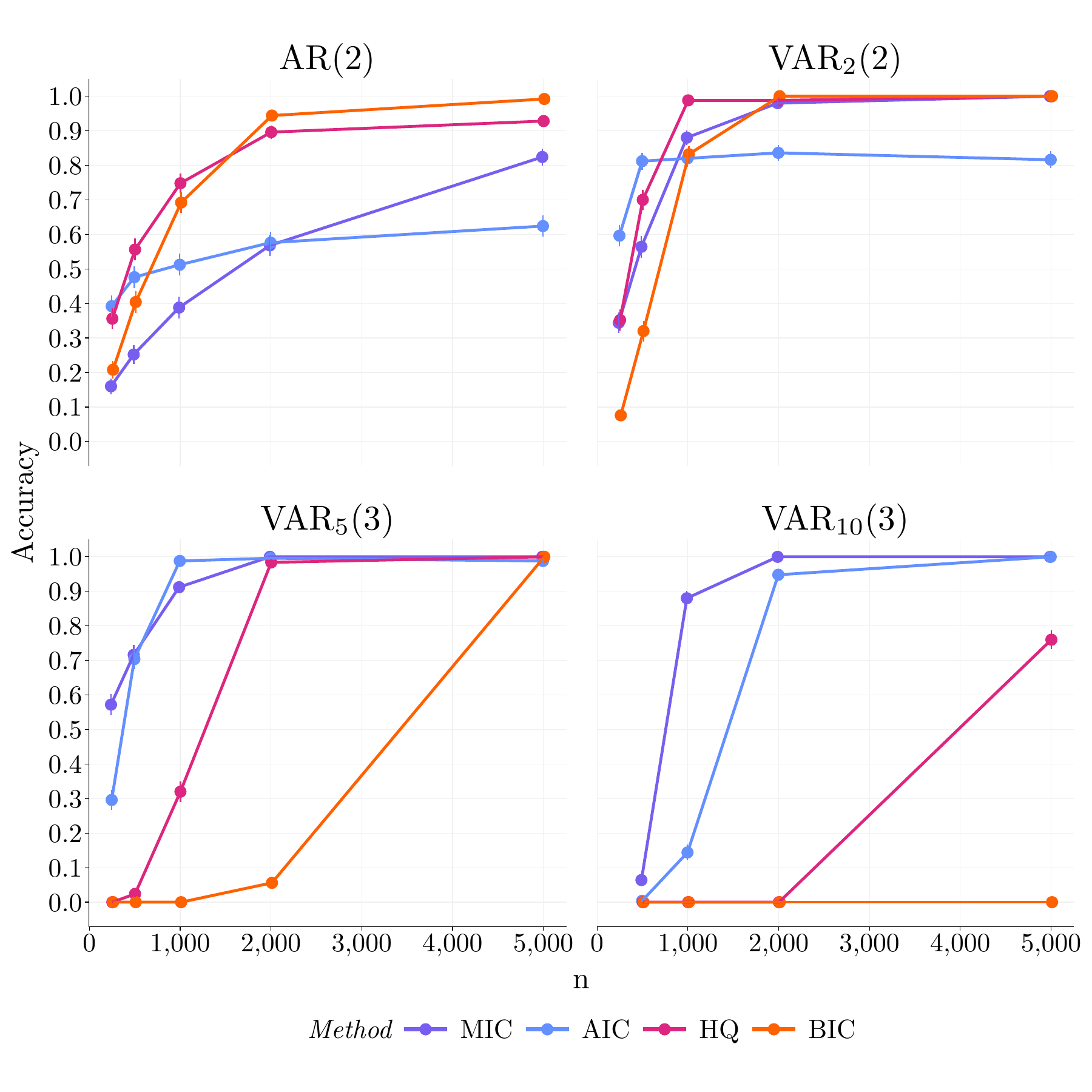}
\caption[Diagonal Gaussian errors simulation results]{\textbf{Diagonal Gaussian errors}. Simulation results for accuracy of specific order selection method and simulation setting with diagonal Gaussian errors. Vertical lines indicate standard errors.}
\label{fig:diagGaus_sims}
\end{figure}

\begin{figure}
\centering
\includegraphics[width=0.95\linewidth]{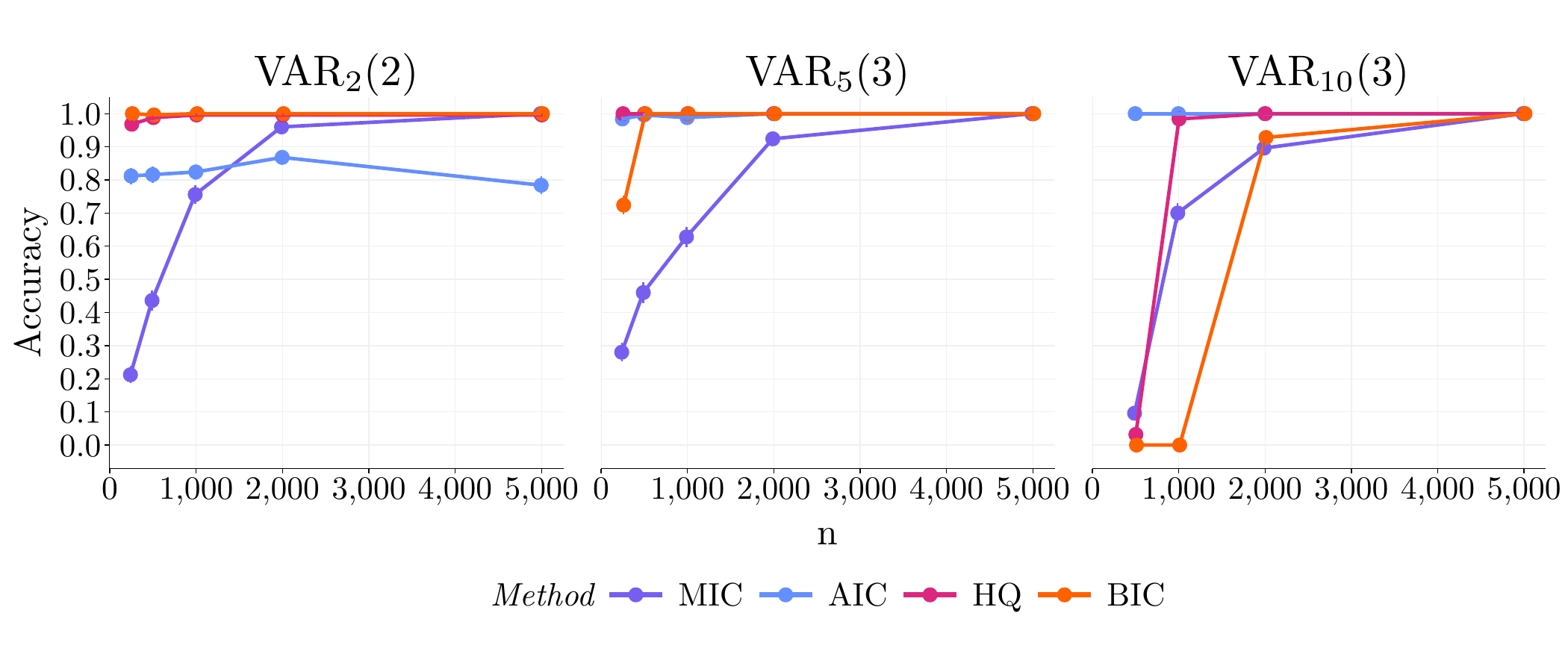}
\caption[Non-diagonal Gaussian errors simulation results]{\textbf{Non-diagonal Gaussian errors}. Simulation results for accuracy of specific order selection method and simulation setting with non-diagonal Gaussian errors. Vertical lines indicate standard errors.}
\label{fig:nonDiagGaus_sims}
\end{figure}

\begin{figure}
\centering
\includegraphics[width=0.95\linewidth]{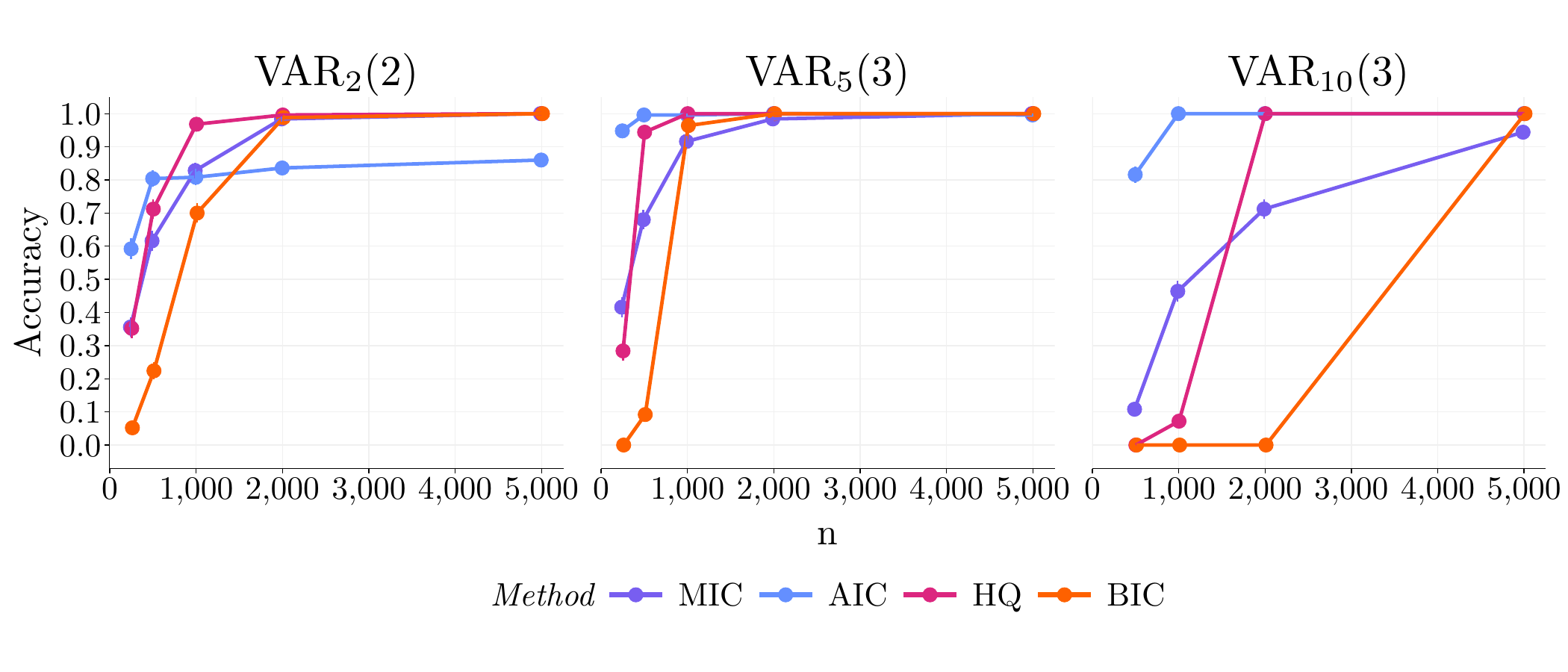}
\caption[Gaussian mixture errors simulation results]{\textbf{Gaussian mixture errors}. Simulation results for accuracy of specific order selection method and simulation setting with Gaussian mixture errors. Vertical lines indicate standard errors.}
\label{fig:gausMix_sims}
\end{figure}

\begin{figure}
    \centering
\includegraphics[width=0.4\linewidth]{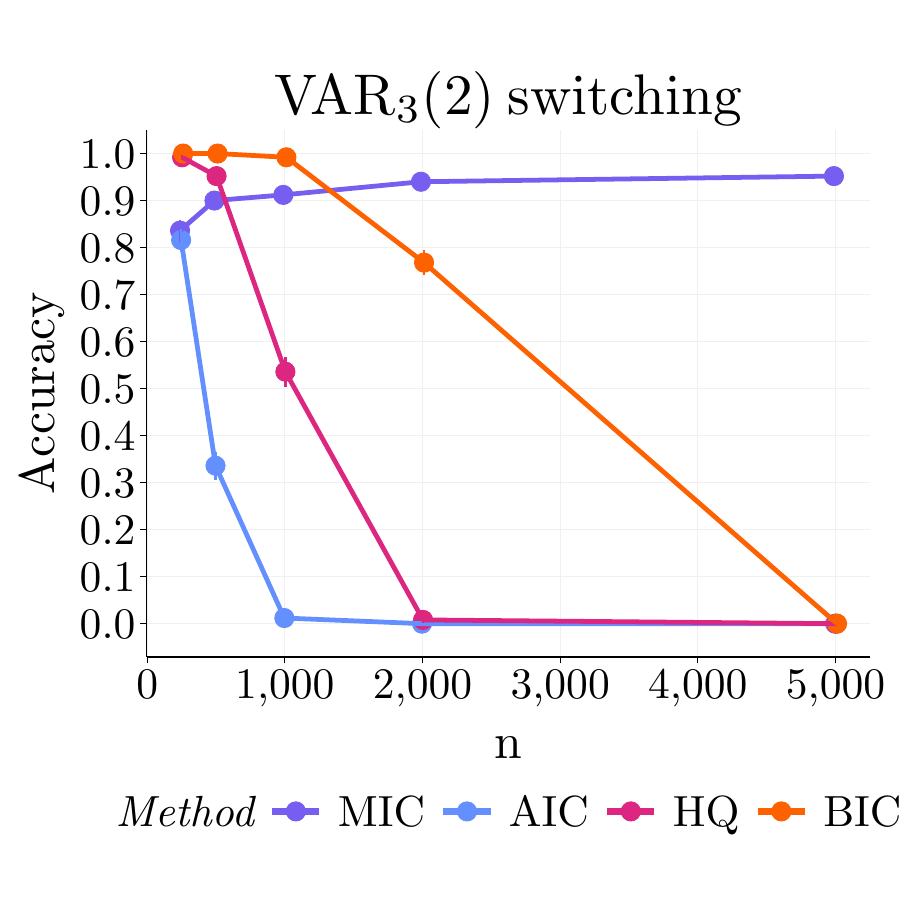}
\caption[$\VAR_3(2)$ switching simulation results]{\textbf{VAR}$\mathbf{_3(2)}$\textbf{ switching}. Simulation results for accuracy of specific order selection method and simulation setting. Vertical lines indicate standard errors.}
    \label{fig:var3_2_switch_sims}
\end{figure}

We compute each criteria, MIC, AIC, BIC, and HQ, for $p = 0, \dots, 10 := p_{\max}$. The estimated orders for each criteria are those that achieve the minimum value. 
That is,
\[
\hat{p}_{\mathrm{CRITERION}} = \argmin_{p \in \{0, \dots, p_{\max}\}} \mathrm{CRITERION}(p) \, .\\
\]
For each error structure and VAR model, we simulate $B = 250$ data sets and compute the proportion of times the correct order is estimated. That is, we use
\[
\text{Accuracy} = \frac{1}{B} \sum_{b=1}^{B} I(\hat{p}_{\mathrm{CRITERION}} = p_0) \, .
\]
The simulation results are summarized in \cref{fig:diagGaus_sims,fig:nonDiagGaus_sims,fig:gausMix_sims,fig:var3_2_switch_sims}. When errors are diagonal Gaussian and the dimension is large, MIC outperforms AIC, HQ, and BIC, as shown in \cref{fig:diagGaus_sims}. This makes sense as in this setting AIC, HQ, and BIC all use the entire error matrix, including the off-diagonals, through the determinant. When the true errors are diagonal, the estimated off-diagonals can contain incorrect information. On the other hand, MIC only uses the diagonals and so discards the potentially misleading off-diagonal information. For diagonal errors with small sample sizes and small dimension, HQ, and BIC perform slightly better, but MIC is still competitive. As previously noted, AIC is not consistent for the $\mathrm{AR}(2)$ and $\VAR_2(2)$ processes. Results for non-diagonal Gaussian errors in \cref{fig:nonDiagGaus_sims} show much better performance for AIC, BIC, and HQ. MIC still appears to consistently estimate the order as sample size increases, but the small sample performance is now worse relative to the other methods. This makes sense as there is useful information contained in the off-diagonals of the error matrix that AIC, BIC, and HQ can leverage while MIC does not. It is also worth noting that the performance of MIC relative to other methods improves as the dimension of the process increases. We see similar trends for Gaussian mixture errors in \cref{fig:gausMix_sims}. Note that performance is sometimes better and sometimes worse than the corresponding results in \cref{fig:nonDiagGaus_sims}. This is likely due to variation in the simulated error covariances and means.

For the misspecification simulation---the $\VAR_{3}(2)$ switching process in \cref{fig:var3_2_switch_sims}---the performance of AIC, BIC, and HQ deteriorates as sample size increases. While the performance of AIC deteriorates fastest, all methods except MIC have an accuracy of 0 for $n = 5000$. Conversely, the accuracy of MIC increases a $n$ increases suggesting it consistently estimates the true order. To investigate this further, we explore the performance of each method with much larger sample sizes in \cref{fig:var3_2_switch_large_sims} and find that the results are consistent with \cref{fig:var3_2_switch_sims}. That is, AIC, BIC, and HQ all have 0 accuracy for $n > 20000$. In contrast, $\mathrm{MIC}$ shows robust performance and perfectly estimates the order for all sample sizes in \cref{fig:var3_2_switch_large_sims}.

While MIC is outperformed by AIC, BIC, and HQ when the true error matrix has off-diagonal elements, it offers better performance when the true error matrix is diagonal and appears to be more robust to misspecification as shown by the performance in the $\VAR_3(2)$ switching setting. Thus,  $\mathrm{MIC}$ offers a viable alternative to AIC/BIC/HQ, as in many practical applications errors are unlikely to be Gaussian. In \cref{sec:applications}, we investigate the performance of MIC compared to AIC/BIC/HQ by comparing forecast performance on two datasets.

\subsection{Over and under selection probability}
\label{sec:over_under_selection}
As previously mentioned, it is known that AIC does not provide a consistent estimate of the VAR order when the dimension is small. To investigate this, we compare the likelihood of over and under selection of the model order for each of the order selection methods using simulations. Since MIC uses an estimated $\lambda$, we evaluate its theoretical properties with an oracle $\lambda$ value, which we denote as MIC-oracle. Specifically, we choose $\lambda_{\mathrm{oracle}} = M/2$ where $M$ is defined in \cref{cor:param_defn}. The results for diagonal Gaussian errors are shown in \cref{fig:over_under_diagGaus_sims1,fig:over_under_diagGaus_sims2} while the non-diagonal gaussian errors are shown in \cref{fig:over_under_nonDiagGaus_sims}. Overall, we see that MIC-oracle tends to select an order that is larger than the true order (over selection) while the alternative methods AIC, BIC, and HQ, tend to select an order that is smaller than the true order (under selection). \cref{fig:over_under_nonDiagGaus_sims} shows that, when the dimension of the process is large and errors are non-diagonal Gaussian, MIC-oracle suffers from worse over selection relative to the under selection from AIC, BIC, and HQ.

\section{Applications}\label{sec:applications}

In this section, we apply our MIC method to two different forecasting problems and compare it to AIC, BIC, and HQ. The first problem is financial forecasting, while the second is forecasting COVID-19 outcomes. In both problems, we follow \citet{nicholson2020high} by comparing the weighted mean squared forecast error (wMSFE). In both problems, the first 80\% of observations are used to estimate the order for MIC, AIC, BIC, and HQ, while the last 20\% are used for evaluating the forecast accuracy. Formally, if $n$ represents the total number of observations, then the first $T_1 = \lfloor 0.8n \rfloor$ data points are used to estimate the order for each method and the remaining observations are used for testing. We use a rolling window of size $T_1$ to perform one-step ahead forecasts. That is, if $t$ indexes the observation we are forecasting, then we use observations ${t - T_1}, \dots, {t - 1}$ as the rolling window. For each rolling window, we standardize each variable in the series by subtracting the mean and dividing by the standard deviation. The observation we are forecasting is also standardized using the mean/SD from the rolling window. We then fit a VAR model corresponding to the orders chosen by MIC, AIC, BIC, and HQ to this standardized rolling window and predict observation $t$. The wMSFE for method $m$ is computed over all series and forecast time points as
\begin{equation*}
\mathrm{wMSFE}(m) = \frac{1}{k(n - T_1)} \sum_{i=1}^{k} \sum_{t=T_1 + 1}^{n} \left( \frac{y_{i,t} - \hat{y}^m_{i,t}}{\hat{\sigma}_i} \right)^2 \, ,
\end{equation*}
where $\hat{\sigma}_i$ is the standard deviation of the variable $i$ computed over the forecast observations.

\subsection{Daily realized stock variances}
\label{sec:stock_analysis}

We compare the forecast performance of VAR models with order selected by MIC, AIC, BIC, and HQ using data from the Oxford-Man Institute of Quantitative Finance obtained from an older version of the \texttt{mfGARCH} \texttt{R} package, \href{https://github.com/onnokleen/mfGARCH/tree/3d6228a8bb861ac0c37cf941f1d4131c99f3e597}{https://github.com/onnokleen/mfGARCH}, from \citet{conrad2020two}. Specifically, we analyze 5-minute return daily realized variances for up to 17 stocks from January 3, 2000 to June 27, 2018 ($n = 4,847$). We perform two analyses: one using the same $k=16$ stocks as in \citet{nicholson2020high} as well as $k=7$ stocks from \citet{son2023forecasting}. Many stocks from  \citet{nicholson2020high} and \citet{son2023forecasting} overlap and there are only 17 total unique stocks. Due to high levels of missingness (34\%) we exclude OMXSPI, an index of the Stockholm Stock Exchange, from our analysis based on the \citet{son2023forecasting} stocks. A full list of the stocks analyzed is given in \cref{app:finance_application}. All data are log-transformed to make them stationary. As we are not specifically interested in high-dimensional applications we estimate the order for each order selection method using a $p_{\max} = 10$, equivalent to two trading weeks.

Forecast results for both the $k = 16$ and $k = 7$ analysis are displayed in \cref{tbl:wmsfe_finance}. Overall, we see that the methods considered give very similar forecast accuracy despite a large range of orders. For example, in \cref{tbl:wmsfe_finance}(a), the order varies from a low of 2 chosen by BIC to a high of 9 chosen by MIC.

\begin{table}[!htb]
\centering
\begin{tabular}[t]{cccccc}
\toprule
  &   & AIC & BIC & HQ & MIC\\
\midrule
 & Order & 7 & 2 & 4 & 9\\
\multirow{-2}{*}[0.5\dimexpr\aboverulesep+\belowrulesep+\cmidrulewidth]{\centering\arraybackslash $k = 16$} & Forecast Error & 0.485 & 0.504 & 0.487 & 0.491\\
\cmidrule{1-6}
 & Order & 8 & 4 & 6 & 6\\
\multirow{-2}{*}[0.5\dimexpr\aboverulesep+\belowrulesep+\cmidrulewidth]{\centering\arraybackslash $k=7$} & Forecast Error & 0.493 & 0.497 & 0.492 & 0.492\\
\bottomrule
\end{tabular}
\caption[Stock variance forecasting]{Comparison of order selection methods based on weighted Mean Squared Forecast Error for daily realized stock variances for $k = 16$ stocks and $k = 7$ stocks. Order selected by each method is also included.}
\label{tbl:wmsfe_finance}
\end{table}

\subsection{COVID-19 in New York City}

We next compare the performance of order selection methods in forecasting COVID-19 outcomes in New York City. Daily data on deaths, cases, and hospitalizations in New York City due to COVID-19 are available starting February 29, 2020 at City of New York's \href{https://data.cityofnewyork.us/Health/COVID-19-Daily-Counts-of-Cases-Hospitalizations-an/rc75-m7u3/about\_data}{website}. We analyze data from February 29, 2020 to July 8, 2024 ($n = 1,592$). All data are first differenced to make them stationary and we use $p_{\max} = 30$. As a check, we run an Augmented Dickey-Full test  \citep[ADF,][]{said1984testing} and a Kwiatkowski-Phillips-Schmidt-Shin  \citep[KPSS,][]{kwiatkowski1992testing} test for each series after differencing. The null hypothesis of the ADF test is that a unit root is present in the time series while the null hypothesis of the KPSS test is that the series is trend-stationary. All series pass the ADF test with $p < 0.01$ and the KPSS test with $p > 0.1$.

Forecast results are displayed in \cref{tbl:wmsfe_covid}. We see that AIC, BIC, and HQ all fit models using around a month's worth of prior data points ($p = 30$) to forecast the next day. However, these models are substantially worse than the model fitted using MIC order selection, which only uses around a week of prior data points ($p = 8$). In fact, the forecast accuracy of the model fitted using MIC is around 30\% better than the accruacy of those fit by AIC/BIC/HQ.

\begin{table}[!htb]
\centering
\begin{tabular}[t]{ccccc}
\toprule
   & AIC & BIC & HQ & MIC\\
\midrule
 Order & 30 & 24 & 30 & 8\\
 Forecast Error & 1.334 & 1.301 & 1.334 & 1.036\\
\bottomrule
\end{tabular}
\caption[COVID-19 forecasting]{Comparison of order selection methods based on weighted Mean Squared Forecast Error for COVID-19 outcomes. Order selected by each method is also included.}
\label{tbl:wmsfe_covid}
\end{table}

\section{Discussion}\label{sec:discussion}

In this paper, we proposed the mean square information criterion (MIC), a new approach for estimating the order of VAR processes. MIC is based on a key new observation: the flatness of the expected squared error loss after the fitted order exceeds the true order. We show, under relatively mild assumptions, that the true order can be estimated consistently by minimizing the MIC. Specifically, consistency of MIC only requires consistent estimates of the autocovariances.

Our proposed method, $\mathrm{MIC}$, was compared to three other order selection criteria, AIC, BIC, and HQ, based on the proportion of simulations in which the correct order was correctly estimated. Simulation settings ranged from univariate to 10-dimensional VAR models with between 250 to 5,000 observations. Simulations included both Gaussian and Gaussian mixture error structures, as well as a regime switching $\VAR_3(2)$ model. While outperformed in Gaussian errors and small sample sizes, relative to the other criteria, $\mathrm{MIC}$ showed the best performance when the process had regime changes. As errors are unlikely ever Gaussian, these results suggest that $\mathrm{MIC}$ can be very useful in practice. This is confirmed in our data applications where order selection via MIC achieved comparable forecast accuracy for daily realized stock variance and substantially better accuracy in forecasting COVID-19 outcomes in NYC when compared to order selected via AIC, BIC, or HQ.

An interesting direction for future work is to extend the proposed method to high-dimensional settings. In high dimensions, we can substitute the least squares estimate by e.g. ridge or LASSO estimators.  Alternatively, estimates of the autocovariances in high dimensions may be used and plugged directly into the loss. It would then be interesting to compare the resulting estimator to recently proposed regularization-based approaches \citep{shojaie2010discovering, shojaie2012adaptive, nicholson2017varx}.

\bibliographystyle{plainnat}
\bibliography{references}
\clearpage
\appendix
\onecolumn
\addcontentsline{toc}{section}{Appendix} 
\part{Appendix} %
\parttoc %

\clearpage

\renewcommand{\thetable}{S\arabic{table}}
\renewcommand{\thefigure}{S\arabic{figure}}%

\section{Assumptions} \label{app:assumptions}
In this section, we list the assumptions used in our analyses and provide a brief discussion of each. All these assumptions are mild.

\begin{assumption}\label{asm:stability}
    $Z_t \in \mathbb{R}^{k \times 1}$ is a stable mean zero process. That is, $\det(I_k - A_1 z - \dots - A_{p_0} z^{p_0}) \neq 0$ for $|z| \leq 1$.
\end{assumption}

This stability condition is standard and is identical to that used in \citet[Eq. (2.1.12)]{lutkepohl2005new}. Note that stability implies stationarity \citep[Proposition 2.1]{lutkepohl2005new} and this assumption is required to replace second order expectations with autocovariances. That is, \cref{asm:stability} is required to have $\E{Z_{t} Z_{t-h}^T} = \Gamma_{h}$. It is also required to use the Yule-Walker equations.

\begin{assumption}\label{asm:invertibility}
    \begin{equation*}
        \E{X_{p_{\max}} X_{p_{\max}}^T } = \begin{bmatrix}
        \Gamma_0 & \Gamma_1 & \dots & \Gamma_{p_{\max}-1} \\
        \Gamma_1^T & \Gamma_0 & \dots & \Gamma_{p_{\max}-2} \\
        \vdots & & & \vdots \\
        \Gamma_{p_{\max}-1}^T & \Gamma_{p_{\max}-2}^T & \dots & \Gamma_0
    \end{bmatrix} \, ,
    \end{equation*} is invertible.
\end{assumption}

Note that due to the quadratic form of $\E{X_{p_{\max}} X_{p_{\max}}^T } $, this matrix is symmetric and positive semidefinite. By assuming invertibility, we ensure this matrix is also positive definite. We also only need to make this assumption for $p_{\max}$ and we will have positive definiteness for all $\E{X_{p_{\max}} X_{p_{\max}}^T } $ for $i = 1, \dots p_{\max}$ since a matrix is positive definite if and only if all its principle minors are positive (see \cite{lutkepohl2005new} Appendix A.8.3). The $ki^{th}$ principle minor of $\E{X_{p_{\max}} X_{p_{\max}}^T } $ is $\det \left(\E{X_{i} X_{i}^T} \right)$. Again, $\E{X_{i} X_{i}^T}$ is symmetric and positive semi-definite so ensuring a positive determinant ensures strictly positive eigenvalues and thus positive definiteness. In all simulation settings this assumption has been met.

\begin{assumption}\label{asm:irreducibility}
    We assume that for a VAR($p_0$) process when $p < p_0$,  
    \[
    \begin{bmatrix} \Gamma_1 & \dots & \Gamma_{p-1} \end{bmatrix}  
    \begin{bmatrix} \Gamma_0 & \dots & \Gamma_{p-1} \\ 
                                \vdots & & \vdots \\
                                \Gamma_{p-1}^T & \dots & \Gamma_0 
    \end{bmatrix}^{-1}  
    \begin{bmatrix} \Gamma_1^T \\ \vdots \\ \Gamma_p^T \end{bmatrix} - 
    \Gamma_p \neq 0. 
    \]
\end{assumption}

This assumption essentially states that we need at least $p_0$ lags to generate the $p{\mathrm{th}}$ autocovariance $\Gamma_p$.  This is a mild assumption. For example, it is implicitly made in \citet[Eq.~(2.1.37)]{lutkepohl2005new} where for a VAR($p_0$) process, the autocovariance matrix is determined by the prior $p_0$ lags.

\section{Simplifying multivariate loss}\label{app:simplify_loss}

In this section, we show that the profiled loss can be written as
\begin{equation*}
    \Eb{\left(Y_t - A_p^*X_p \right)^T\left(Y_t - A_p^* X_p \right)} = \Tr\left( \Gamma_0 \right) - 
    \Tr \left( 
    \begin{bmatrix} \Gamma_1 & \dots & \Gamma_p \end{bmatrix}  
    \begin{bmatrix} \Gamma_0 & \dots & \Gamma_{p-1} \\ 
                                    \vdots & & \vdots \\
                                    \Gamma_{p-1}^T & \dots & \Gamma_0 
    \end{bmatrix}^{-1}  
    \begin{bmatrix} \Gamma_1^T \\ \vdots \\ \Gamma_p^T \end{bmatrix}
    \right) \, .
\end{equation*}
To write the profiled loss in this way, we will need to replace expectations with atuocovariances as in $\E{Z_{t} Z_{t-h}^T} = \Gamma_{h}$. Thus \cref{asm:stability} (stability) is required.

\subsection{Simplifying equalities}

To begin, we first compute some expectations that will be needed to simplify the loss. We make note of several identities we will use. First, $\text{vec}(A)^T\text{vec}(B) = \Tr(A^TB)$ and $\text{vec}(ABC) = (C^T \otimes A)\text{vec}(B)$. The $j^{\mathrm{th}}$ row and column of $\Gamma_{i}$ are denoted as $\Gamma_{i,j\cdot}$ and $\Gamma_{i,\cdot j}$ respectively.

We also note that $\E{X_p^T \otimes X_p^T} \in \mathbb{R}^{ 1 \times k^2p^2}$ and can be written as
\begin{equation*}
    \begin{split}
        \E{X_p^T \otimes X_p^T} & = \E{\begin{bmatrix}
        Z_{t-1,1} & \dots & Z_{t-1,k} & Z_{t-2,1} & \dots & Z_{t-p,k} \end{bmatrix} \otimes \begin{bmatrix}
        Z_{t-1,1} & \dots & Z_{t-1,k} & Z_{t-2,1} & \dots & Z_{t-p,k} \end{bmatrix}} \\
        & = \E{\begin{matrix}
            [\text{ } Z_{t-1,1}^2 & \dots & Z_{t-1,1}Z_{t-1,k} & Z_{t-1,1}Z_{t-2,1} & \dots & Z_{t-1,1}Z_{t-2,k} & \dots & Z_{t-1,1} Z_{t-p,k} \\
             Z_{t-1,2}Z_{t-1,1} & \dots & Z_{t-1,2}Z_{t-1,k} & Z_{t-1,2}Z_{t-2,1} & \dots & Z_{t-1,2}Z_{t-2,k} & \dots & Z_{t-1,2} Z_{t-p,k} \\
             \vdots &  & \vdots & \vdots  & & \vdots &  & \vdots  \\
             Z_{t-1,k}Z_{t-1,1} & \dots & Z_{t-1,k}Z_{t-1,k} & Z_{t-1,k}Z_{t-2,1} & \dots & Z_{t-1,k}Z_{t-2,k} & \dots & Z_{t-1,k} Z_{t-p,k}\\
             Z_{t-2,1}Z_{t-1,1} & \dots & Z_{t-2,1}Z_{t-1,k} & Z_{t-2,1}Z_{t-2,1} & \dots & Z_{t-2,1}Z_{t-2,k} & \dots & Z_{t-2,1} Z_{t-p,k} \\
             \vdots &  & \vdots & \vdots  & & \vdots &  & \vdots \text{ }] \\
        \end{matrix}} .
    \end{split}
\end{equation*}
We have 
\begin{equation*}
    \E{\begin{bmatrix} Z_{t-1,1}^2 & \dots & Z_{t-1,1}Z_{t-1,k} \end{bmatrix}^T } = \Gamma_{0, \cdot 1} .
\end{equation*}
Similarly, 
\begin{align*}
    \E{\begin{bmatrix}Z_{t-1,1}Z_{t-2,1} & \dots & Z_{t-1,1}Z_{t-2,k}\end{bmatrix}^T} & = \Gamma_{-1, \cdot 1}  = (\Gamma_{1}^T)_{\cdot 1} \, , \\
    \E{\begin{bmatrix}Z_{t-1,1}Z_{t-p,1} & \dots & Z_{t-1,1}Z_{t-p,k}\end{bmatrix}^T} & = \left(\Gamma_{p-1}^T\right)_{\cdot 1} \, .
\end{align*}
We define
\begin{equation*}
\begin{split}
 C & := \begin{matrix}
    [ \Gamma_{0, \cdot 1} & (\Gamma_{1}^T)_{\cdot 1} &  (\Gamma_{2}^T)_{ \cdot 1} & \dots & (\Gamma_{p-1}^T)_{\cdot 1} \\
     \Gamma_{0, \cdot 2} & (\Gamma_{1}^T)_{\cdot 2} & (\Gamma_{2}^T)_{\cdot 2} & \dots & (\Gamma_{p-1}^T)_{\cdot 2} \\
     \vdots & & &  & \vdots \\
     \Gamma_{0, \cdot k} & (\Gamma_{1}^T)_{\cdot k} & (\Gamma_{2}^T)_{\cdot k} & \dots & (\Gamma_{p-1}^T)_{\cdot k} \\
     \Gamma_{1, \cdot 1} & \Gamma_{0, \cdot 1} & (\Gamma_{1}^T)_{\cdot 1} & \dots & (\Gamma_{p-1}^T)_{\cdot k} \\
     \vdots & & &  & \vdots \\ 
     \Gamma_{p-1, \cdot k} & \Gamma_{p-2, \cdot k} & \Gamma_{p-3, \cdot k} & \dots & \Gamma_{0, \cdot k} ] \, .
\end{matrix} 
\end{split}
\end{equation*}
Note that each element in C is $\in \mathbb{R}^{k \times 1}$. For example, both $\Gamma_{0, \cdot 1}$ and $(\Gamma_{1})^T_{\cdot 1}$ are $\in \mathbb{R}^{k \times 1}$. Thus, $C \in \mathbb{R}^{k \times kp^2}$, and we can write

\begin{equation*}
\begin{split}
\E{X_p^T \otimes X_p^T} & = \mathrm{vec}(C)^T \\
& = \text{vec}\left(\begin{bmatrix}
    \Gamma_0 & \Gamma_1 & \dots & \Gamma_{p-1} \\
    \Gamma_1^T & \Gamma_0 & \dots & \Gamma_{p-2} \\
    \vdots & & & \vdots \\
    \Gamma_{p-1}^T & \Gamma_{p-2}^T & \dots & \Gamma_0
\end{bmatrix} \right)^T \, .
\end{split}
\end{equation*}

Using a similar idea, we have that
\begin{equation*}
\begin{split}
\E{X_p^T \otimes Y_t^T} & = \E{\begin{bmatrix}
Z_{t-1,1} & \dots & Z_{t-1,k} & Z_{t-2,1} & \dots & Z_{t-p,k} \end{bmatrix} \otimes \begin{bmatrix}
Z_{t,1} & \dots & Z_{t,k}  \end{bmatrix}} \\
& = \text{vec}\left( \begin{bmatrix} \Gamma_1 & \dots & \Gamma_{p} \end{bmatrix} \right)^T \, .
\end{split}
\end{equation*}
For space considerations, we do not write out the steps in detail. The method proceeds similarly to the above. Next,
\begin{equation*}
    \begin{split}
        \E{X_p X_p^T}  = \E{\begin{bmatrix}
            Z_{t-1} \\
            Z_{t-2} \\
            \vdots \\
            Z_{t-p}
        \end{bmatrix} \begin{bmatrix} Z_{t-1}^T & Z_{t-2}^T & \dots & Z_{t-p}^T \end{bmatrix}} 
        & = \E{\begin{bmatrix} Z_{t-1}Z_{t-1}^T & Z_{t-1}Z_{t-2}^T & \dots & Z_{t-1}Z_{t-p}^T \\
        Z_{t-2}Z_{t-1}^T & Z_{t-2}Z_{t-2}^T & \dots & Z_{t-2}Z_{t-2}^T \\
        \vdots & & &  \vdots \\
        Z_{t-p}Z_{t-1}^T & Z_{t-p}Z_{t-2}^T & \dots & Z_{t-p}Z_{t-p}^T
        \end{bmatrix}} \\
        & = \begin{bmatrix}
            \Gamma_0 & \Gamma_1 & \dots & \Gamma_{p-1} \\
            \Gamma_1^T & \Gamma_0 & \dots & \Gamma_{p-2} \\
            \vdots & & & \vdots \\
            \Gamma_{p-1}^T & \Gamma_{p-2}^T & \dots & \Gamma_0 
        \end{bmatrix} \, .
    \end{split}
\end{equation*}
Lastly,
\begin{equation*}
    \E{Y_t X_p^T} = \begin{bmatrix} \Gamma_1 & \dots & \Gamma_p \end{bmatrix} \, .
\end{equation*}

\subsection{Simplifying the loss}

We can now proceed in simplifying the population loss. First, note that 
\begin{equation*}
\begin{split}
    \Eb{\left(Y_t - A_p^*X_p \right)^T\left(Y_t - A_p^*X_p \right)}  & = \E{Y_t^TY_t} - 2\E{ Y_t^T A^*_p X_p} + \E{X_p^T A^{* T}_p A^{*}_p X_p} \\
    & = \Tr\left(\Gamma_0\right) - 2 \E{X_p^T \otimes Y_t^T} \text{vec}\left( A^{*}_p \right) + \E{X_p^T \otimes X_p^T} \text{vec}\left( A^{* T}_p A^{*}_p \right) \\
    & = \Tr\left( \Gamma_0 \right) - 2 \Tr \left( \begin{bmatrix} \Gamma_1^T \\ \vdots \\ \Gamma_p^T \end{bmatrix} A^*_p \right) + \Tr\left( \begin{bmatrix} \Gamma_0 & \dots & \Gamma_{p-1} \\ 
                                \vdots & & \vdots \\
                                \Gamma_{p-1}^T & \dots & \Gamma_0 
                \end{bmatrix} A^{* T}_p A^{*}_p\right) \, .
\end{split}
\end{equation*}

Now, $A^{* T}_p =  \E{X_p X_p^T}^{-1,T}E\left( Y_t X_p^T \right)^{T} = \begin{bmatrix} \Gamma_0 & \dots & \Gamma_{p-1} \\ 
                                \vdots & & \vdots \\
                                \Gamma_{p-1}^T & \dots & \Gamma_0 
                \end{bmatrix}^{-1,T} \begin{bmatrix} \Gamma_1^T \\ \vdots \\ \Gamma_p^T \end{bmatrix}  $. Thus, $\begin{bmatrix} \Gamma_0 & \dots & \Gamma_{p-1} \\ 
                                \vdots & & \vdots \\
                                \Gamma_{p-1}^T & \dots & \Gamma_0 
                \end{bmatrix}$ is symmetric and so is its inverse. Therefore, using $\Tr(A^TB) = \Tr(AB^T)$, we get

\begin{equation*}
\begin{split}
 \Eb{\left(Y_t - A_p^{*}X_p \right)^T\left(Y_t - A_p^{*}X_p \right)}  & = \Tr\left( \Gamma_0 \right) - \Tr \left( \begin{bmatrix} \Gamma_1^T \\ \vdots \\ \Gamma_p^T \end{bmatrix} A^*_p \right) \\
& =  \Tr\left( \Gamma_0 \right) - 
\Tr \left( 
\begin{bmatrix} \Gamma_1 & \dots & \Gamma_p \end{bmatrix}  
\begin{bmatrix} \Gamma_0 & \dots & \Gamma_{p-1} \\ 
                                \vdots & & \vdots \\
                                \Gamma_{p-1}^T & \dots & \Gamma_0 
\end{bmatrix}^{-1}  
\begin{bmatrix} \Gamma_1^T \\ \vdots \\ \Gamma_p^T \end{bmatrix}
\right) \, .
\end{split}
\end{equation*}

\section{Proofs}

In this section, we prove \cref{thm:flat_loss,cor:param_defn,thm:consistency}. Specifically, \cref{app:flat_loss} focuses on \cref{thm:flat_loss}, \cref{app:recover_true_order} proves \cref{cor:param_defn}, and \cref{app:consistent_sample_loss} proves \cref{thm:consistency}.

\subsection{Flat loss} \label{app:flat_loss}

In this section, we prove \cref{thm:flat_loss} which establishes that the loss decreases until the true order $p_0$ at which point it remains constant at $\Tr(\Sigma_{\epsilon})$. The proof proceeds in several steps. We first relate the loss at fitted order $p$ to the loss at fitted order $p - 1$. In the second step, we consider the cases when $p > p_0$ and $p \leq p_0$ separately. Finally, we show that the loss at $p_0$ is equal to $\Tr(\Sigma_{\epsilon})$.

\begin{proof}[Proof of \cref{thm:flat_loss}]

Recall that
\begin{equation*}
    \LVAR(p) := \Tr\left( \Gamma_0 \right) - \Tr \left( 
        \begin{bmatrix} \Gamma_1 & \dots & \Gamma_p \end{bmatrix}  
        \begin{bmatrix} \Gamma_0 & \dots & \Gamma_{p-1} \\ 
                                        \vdots & & \vdots \\
                                        \Gamma_{p-1}^T & \dots & \Gamma_0 
        \end{bmatrix}^{-1}  
        \begin{bmatrix} \Gamma_1^T \\ \vdots \\ \Gamma_p^T \end{bmatrix}
    \right) \, .
\end{equation*}
To express $\LVAR(p)$ as a function of $\LVAR(p-1)$, we partition the matrices as follows
\begin{equation*} 
    \left[
    \begin{array}{ccc|c}
    \Gamma_1 & \dots & \Gamma_{p-1} & \Gamma_p
    \end{array} \right] := 
    \begin{bmatrix}
        g & \Gamma_p
    \end{bmatrix} \, , 
\end{equation*}
\begin{equation} 
\label{eqn:partition_mat}
    \left[
        \begin{array}{ccc|c}
        \Gamma_0 & \dots & \Gamma_{p-2} & \Gamma_{p-1} \\
        \Gamma_1^T & \dots & \Gamma_{p-3} & \Gamma_{p-2} \\
        \vdots & & \vdots & \vdots \\
        \Gamma_{p-2}^T & \dots & \Gamma_{0} & \vdots \\
        \hline 
        \Gamma_{p-1}^T & \dots & \Gamma_{1}^T & \Gamma_0
        \end{array} 
    \right] := 
    \begin{bmatrix}
        B & C^T \\
        C & D
    \end{bmatrix} \, .
\end{equation}
Now, using from the 2 x 2 block matrix inversion formula \citep{lu2002inverses},
\begin{equation*}
    \begin{bmatrix} \Gamma_0 & \dots & \Gamma_{p-1} \\ 
                                    \vdots & & \vdots \\
                                    \Gamma_{p-1}^T & \dots & \Gamma_0 
                    \end{bmatrix}^{-1} = 
    \begin{bmatrix}
        B & C^T \\
        C & D
    \end{bmatrix}^{-1} = 
    \begin{bmatrix}
        B^{-1} + B^{-1}C^THCB^{-1} &  -B^{-1}C^TH \\
        -HCB^{-1} & H
    \end{bmatrix} \, ,
\end{equation*}
where $H = (D - CB^{-1}C^{T})^{-1}$ is the inverse of the Schur-complement of block B. Carrying out the multiplication,
\begin{equation*}
    \begin{split}
        &\begin{bmatrix} \Gamma_1 & \dots & \Gamma_p \end{bmatrix}  
        \begin{bmatrix} \Gamma_0 & \dots & \Gamma_{p-1} \\ 
                                        \vdots & & \vdots \\
                                        \Gamma_{p-1}^T & \dots & \Gamma_0 
        \end{bmatrix}^{-1}  
        \begin{bmatrix} \Gamma_1^T \\ \vdots \\ \Gamma_p^T \end{bmatrix} 
        = 
        \begin{bmatrix}
            g & \Gamma_p
        \end{bmatrix}
        \begin{bmatrix}
            B^{-1} + B^{-1}C^THCB^{-1} &  -B^{-1}C^TH \\
            -HCB^{-1} & H
        \end{bmatrix}
        \begin{bmatrix}
            g^T \\ \Gamma_p^T
        \end{bmatrix} \\
     \end{split}
\end{equation*}
\begin{equation*}
     = gB^{-1}g^T + gB^{-1}C^THCB^{-1}g^T - \Gamma_pHCB^{-1}g^{T} - gB^{-1}C^TH\Gamma_p^T + \Gamma_pH\Gamma_p^T \, .
\end{equation*}
Thus,
\begin{equation*}
    \begin{split}
        \LVAR(p) & = \Tr\left(\Gamma_0\right) - \Tr\left( gB^{-1}g^T + gB^{-1}C^THCB^{-1}g^T - \Gamma_pHCB^{-1}g^{T} - gB^{-1}C^TH\Gamma_p^T + \Gamma_pH\Gamma_p^T \right) \\
        & = \LVAR(p-1) - \Tr\left( gB^{-1}C^THCB^{-1}g^T - \Gamma_pHCB^{-1}g^{T} - gB^{-1}C^TH\Gamma_p^T + \Gamma_pH\Gamma_p^T \right) \\
        & = \LVAR(p-1) - \Tr\left(  gB^{-1}C^THCB^{-1}g^T \right) + \Tr\left( \Gamma_pHCB^{-1}g^{T} \right) + \Tr\left( gB^{-1}C^TH\Gamma_p^T  \right) - \Tr\left( \Gamma_pH\Gamma_p^T  \right) \\
        &= \LVAR(p-1) - \Tr\left(\left(g B^{-1} C^T - \Gamma_p \right)H\left(g B^{-1} C^T - \Gamma_p \right)^T \right) \, . 
    \end{split} 
\end{equation*}
Note that this relationship holds in general. To get this result we only rely on \cref{asm:stability} (stability) and \cref{asm:invertibility} (invertibility). 

Now we use that the true process is VAR($p_0$) to study the loss when $p > p_0$. If $p > p_0$, from the Yule-Walker equations,
\begin{equation*}
    \begin{bmatrix} \Gamma_1 & \dots & \Gamma_{p-1} \end{bmatrix} 
    = 
    \begin{bmatrix} A_1 & \dots & A_{p_0} \end{bmatrix} \begin{bmatrix}
        \Gamma_0 & \Gamma_1 & \dots & \Gamma_{p-2} \\
        \Gamma_{-1} & \Gamma_0 & \dots & \Gamma_{p-3} \\
        \vdots & & & \vdots \\
        \Gamma_{-(p_0-1)} & \Gamma_{-(p_0-2)} & \dots & \Gamma_{-(p_0-(p-1))}
    \end{bmatrix}\, .
\end{equation*}
We can extend this to 
\begin{equation*}
    \begin{bmatrix} \Gamma_1 & \dots & \Gamma_{p-1} \end{bmatrix} 
    = 
    \begin{bmatrix} A_1 & \dots & A_{p_0} & 0 & \dots & 0 \end{bmatrix} 
    \begin{bmatrix}
        \Gamma_0 & \Gamma_1 & \dots & \Gamma_{p-2} \\
        \Gamma_{-1} & \Gamma_0 & \dots & \Gamma_{p-3} \\
        \vdots & & & \vdots \\
        \Gamma_{2-p} & \Gamma_{3-p} & \dots & \Gamma_0
    \end{bmatrix}\, .
\end{equation*}
Using that $\Gamma_{-i} = \Gamma_{i}^T$ and substituting in the definitions of $B$ and $g$ and under \cref{asm:invertibility} (invertibility),
\begin{equation*}
    gB^{-1}C^T = \begin{bmatrix} A_1 & \dots & A_{p_0} & 0 & \dots & 0 \end{bmatrix} \begin{bmatrix} \Gamma_{p-1} \\ \vdots \\ \Gamma_1 \end{bmatrix} = \Gamma_p \, .
\end{equation*}
Thus, for any $p > p_0$
\begin{equation*}
    \LVAR(p) = \LVAR(p-1) \, .
\end{equation*}

Next, we study the loss when $p \leq p_0$. From  \cref{asm:invertibility} (invertibility), the matrix
\begin{equation*}
    \begin{bmatrix} \Gamma_0 & \dots & \Gamma_{p-1} \\ 
                                    \vdots & & \vdots \\
                                    \Gamma_{p-1}^T & \dots & \Gamma_0 
    \end{bmatrix} \,
\end{equation*}
is positive definite for all $p = 1, \dots, p_{\max}$. From Proposition 2.2  of \cite{gallier2020schur}, the Schur-complement of block B in \cref{eqn:partition_mat} is also positive definite and hence so is the inverse, $H$. From \cref{asm:irreducibility} (irreducibility), when $p \leq p_0$,
\begin{equation*}
    g B^{-1} C^T - \Gamma_p := \Delta \neq 0 \, .
\end{equation*}
Recall our equation relating $\LVAR(p)$ to $\LVAR(p-1)$:
\begin{equation*}
    \LVAR(p) = \LVAR(p-1) - \Tr\left(\left(g B^{-1} C^T - \Gamma_p \right)H\left(g B^{-1} C^T - \Gamma_p \right)^T \right) \, .
\end{equation*}
Using $\delta_i^T$ to denote the $i^{th}$ row of $\Delta$, 
\begin{equation*}
    \Tr\left(\Delta H \Delta^T \right) = \sum_{i=1}^{k} \delta_i^T H \delta_i \, .
\end{equation*}
Since $H$ is positive definite and $\delta_i \neq 0$ for at least one $i = 1, \dots, k$, $\Tr\left(\Delta H \Delta^T \right) > 0$ and thus
\begin{equation*}
    \LVAR(p) < \LVAR(p-1)\, .
\end{equation*}

Lastly, we show that the loss flattens out at $\Tr(\Sigma_{\epsilon})$. Since $\LVAR(p) = \LVAR(p-1)$ when $p > p_0$, it suffices to show that $\LVAR(p_0) = \Tr(\Sigma_{\epsilon})$:
\begin{equation*}
    \begin{split}
        \LVAR(p_0) &= \Tr\left( \Gamma_0 \right) - 
        \Tr \left( 
        \begin{bmatrix} \Gamma_1 & \dots & \Gamma_{p_0} \end{bmatrix}  
        \begin{bmatrix} \Gamma_0 & \dots & \Gamma_{p_0-1} \\ 
                                        \vdots & & \vdots \\
                                        \Gamma_{p_0-1}^T & \dots & \Gamma_0 
        \end{bmatrix}^{-1}  
        \begin{bmatrix} \Gamma_1^T \\ \vdots \\ \Gamma_{p_0}^T \end{bmatrix}
        \right) \\
        & = \Tr(\Gamma_0) - \Tr\left(\sum_{i=1}^{p_0} A_i \Gamma_{-i}\right) \\
        & = \Tr\left(\Gamma_0 - \left(\Gamma_0 - \Sigma_{\epsilon}\right)\right)\\
        & = \Tr(\Sigma_{\epsilon}) \, ,
    \end{split} 
\end{equation*}
where in the second line we use $\Gamma_i^T = \Gamma_{-i}$, and in the third line we use the form of $\Gamma_0$ from \citet[Eq.~(2.1.36)]{lutkepohl2005new}.

\end{proof}

\subsection{Penalized loss recovers true order} \label{app:recover_true_order}

In this section, we prove \cref{cor:param_defn} which states that, for the correct choice of penalty, the true VAR order can be recovered by penalizing the population loss.

\begin{proof}[Proof of \cref{cor:param_defn}]
    We first show that for $p > p_0$, $\LVAR(p) + \lambda p > \LVAR(p_0) + \lambda p_0$. This follows immediately by noting that $\LVAR(p) - \LVAR(p_0) = 0$ from \cref{thm:flat_loss} and $\lambda p > \lambda p_0$ for $p > p_0$ as $\lambda > 0$. Next, we show that for $p < p_0$, $\LVAR(p) + \lambda p > \LVAR(p_0) + \lambda p_0$. Since $p = p_0 - i$ for some $i$, it suffices to show that $\lambda < \left(\LVAR(p_0 - i) - \LVAR(p_0)\right)/i$ for every $i = 1, \dots, p_0$. This holds by definition of $\lambda$ in \cref{cor:param_defn}.
\end{proof}

\subsection{Consistency of sample loss} \label{app:consistent_sample_loss}

Next, we prove \cref{thm:consistency} which establishes that $\hLVAR(p)$ converges to the population loss $\LVAR(p)$. This is proved using \cref{lem:loss_lipschitz} and the convergence of $\hGamma_i$ to $\Gamma_i$.
\begin{lemma}\label{lem:loss_lipschitz}

Consider the function 
\begin{equation*}
 f(G) = 
\Tr \left( \Gamma_0^0 - 
\begin{bmatrix} \Gamma_1^0 & \dots & \Gamma_p^0 \end{bmatrix}  
\begin{bmatrix} \Gamma_0^1 & \Gamma_1^1 & \Gamma_2^1 &\dots & \Gamma_{p-1}^1 \\ 
                                (\Gamma_1^{1})^{T} & \Gamma_0^{2} & \Gamma_1^{2} & \dots & \Gamma_{p-2}^{2} \\
                                 (\Gamma_2^{1})^{T} & (\Gamma_1^{2})^T & \Gamma_0^{3} & \dots & \Gamma_{p-3}^3 \\
                                \vdots & & & \dots & \vdots\\
                                (\Gamma_{p-1}^{1})^{T} & (\Gamma_{p-2}^{2})^{T} & (\Gamma_{p-3}^{3})^{T} & \dots & \Gamma_0^{p} 
\end{bmatrix}^{-1}  
\begin{bmatrix} (\Gamma_1^{0})^{T} \\ \vdots \\ (\Gamma_p^{0})^{T} \end{bmatrix}
\right) \, ,
\end{equation*}
where the elements of the input vector $G \in \mathbb{R}^{ \frac{k^2(p+2)(p+1)}{2} }$ are denoted as
\begin{equation*}
    G = \begin{bmatrix} \vec{\Gamma}_{0}^{0} & \vec{\Gamma}_1^{0} & \dots & \vec{\Gamma}_p^{0} & \vec{\Gamma}_{0}^{1} & \dots & \vec{\Gamma}_{0}^{p} & \vec{\Gamma}_{1}^{1} & \dots & \vec{\Gamma}_{1}^{p-1} & \vec{\Gamma}_{2}^{1} & \dots & \vec{\Gamma}_{2}^{p-2} & \dots & \vec{\Gamma}_{p-1}^{1}
    \end{bmatrix} \, ,
\end{equation*}
and $\vec{\Gamma} := \mathrm{vec}(\Gamma)^T$ is used to define the row vector formed by stacking the columns of $\Gamma$ and transposing. The subscripts represent the autocovariance lag of interest while the superscripts represent (possibly) different estimates of that autocovariance. That is, $\Gamma_{h}^{i}$ represents the $i^{th}$ estimate of the $h^{th}$ autocovariance.

If $G$ and $\hat{G}$ are defined such that the corresponding inverses in $f(G), f(\hat{G})$ and $f((1-\nu)G + \nu \hat{G})$ exist for all $\nu \in (0,1)$, then $f(G)$ is Lipschitz continuous with respect to $G$. That is
\[
\left| f(\hat{G}) - f(G) \right| \leq L \left\|\hat{G} - G \right\|_{1} \, ,
\]
for some $L < \infty$.
\end{lemma}

The proof of \cref{lem:loss_lipschitz} is deferred until after the proof of \cref{thm:consistency} for clarity of presentation. The proof of \cref{thm:consistency} proceeds by first showing that the least squares estimator of the loss can be expressed in terms of many different autocovariance estimators. Next, we show that the autocovariance estimators in the least squares estimator of the loss are asymptotically equivalent to the autocovariance estimators that use all available data. Finally, this fact is combined with \cref{lem:loss_lipschitz} to establish consistency and the rate of the least squares estimator of the loss.

\begin{proof}[Proof of \cref{thm:consistency}]

Consider our least squares estimator of the loss,
\begin{equation*}
    \begin{split}
        \hLVAR(p) & = \frac{1}{n-p} \Tr\left( \left(\bY_p - \hat{A}_p\mathbf{X}_p \right)\left( \bY_p - \hat{A}_p\mathbf{X}_p \right)^T \right)  \\
        & = \frac{1}{n-p}  \Tr\left( \bY_p \bY_p^T - \bY_p \mathbf{X}_p^T \left(\mathbf{X}_p \mathbf{X}_p^T \right)^{-1}\mathbf{X}_p \bY_p^T \right) \\
        & =  \Tr\left( \frac{1}{n-p} \bY_p \bY_p^T - \left(\frac{1}{n-p} \bY_p \mathbf{X}_p^T \right) \left( \frac{1}{n-p}  \mathbf{X}_p \mathbf{X}_p^T \right)^{-1} \left( \frac{1}{n-p}  \mathbf{X}_p \bY_p^T \right) \right) \, ,
        \end{split} 
\end{equation*}
where for de-meaned data,
\begin{equation*}
    \begin{split}
         \frac{1}{n-p} \bY_p \bY_p^T & =  \frac{1}{n-p}  \sum_{t=1+p}^{n} (\bz_t -\bzbar) (\bz_t - \bzbar)^T \\
         \frac{1}{n-p} \bY_p \mathbf{X}_p^T & = \frac{1}{n-p}  \sum_{t=1+p}^{n} \left(\bz_t - \bzbar \right) \left(\bx_{t, p} - \bzbar \right)^T \\
         & = \frac{1}{n-p}  \begin{bmatrix} \sum_{t=1+p}^{n} \left(\bz_t - \bzbar \right) \left(\bz_{t-1} - \bzbar \right)^T & \dots & \sum_{t=1+p}^{n} \left(\bz_t - \bzbar\right)\left(\bz_{t-p}^T - \bzbar \right) \end{bmatrix} \\
          \frac{1}{n-p} \mathbf{X}_p \mathbf{X}_p^T & = \frac{1}{n-p}  \sum_{t=1+p}^{n} \begin{bmatrix} \left( \bz_{t-1} - \bzbar \right) \\ \vdots \\ \left( \bz_{t-p} -\bzbar \right) \end{bmatrix} \begin{bmatrix} \left( \bz_{t-1} - \bzbar \right)^T & \dots & \left( \bz_{t-p} - \bzbar \right)^ T \end{bmatrix} \\
        & = \frac{1}{n-p}  \sum_{t=1+p}^{n} \begin{bmatrix}
            \left( \bz_{t-1} - \bzbar \right) \left( \bz_{t-1} - \bzbar \right)^T & \left( \bz_{t-1} - \bzbar\right) \left( \bz_{t-2} - \bzbar \right)^T & \dots & \left( \bz_{t-1} - \bzbar \right) \left( \bz_{t-p} - \bzbar \right)^T \\
            \vdots & & & \vdots \\
            \left( \bz_{t-p} - \bzbar \right) \left(\bz_{t-1} - \bzbar \right)^T & \left( \bz_{t-p} - \bzbar \right) \left( \bz_{t-2} - \bzbar \right)^T & \dots & \left( \bz_{t-p}-\bzbar\right) \left(\bz_{t-p}-\bzbar\right)^T 
        \end{bmatrix} \, .
    \end{split} 
\end{equation*}
By examining $ \frac{1}{n-p} \mathbf{X}_p \mathbf{X}_p^T$ we can see that slightly different estimates of the same autocovariance are used. For example, the diagonals of $ \frac{1}{n-p} \mathbf{X}_p \mathbf{X}_p^T$ show that $p$ different estimates of $\Gamma_0$ are used. Similarly $p-1$ different estimates of $\Gamma_1$ are used and $p-2$ different estimates of $\Gamma_2$ are used and so on. All the matrices used in estimating the least squares loss can be written as 
\begin{equation*}
    \hat{\Gamma}_{l-m,p} = \frac{1}{n-p} \sum_{t = 1+p}^{n} (\bz_{t-m} - \bzbar)(\bz_{t-l} - \bzbar)^T \quad \mathrm{for}\ l \in \{0, 1, \dots, p\}, 0 \leq m \leq l \, .
\end{equation*}
These estimators do not use all available data to estimate the autocovariances. For example, to estimate the autocovariance $l - m = 1$, it is possible to use observations $2, \dots, n$. However, the least squares estimator uses only observations $p+1, \dots, n$. We denote the autocovariance estimates that use all available data as
\begin{equation*}
    \tilGamma_{l-m} = \frac{1}{n} \sum_{t=(l-m) +1}^n (\bz_{t} - \bzbar)(\bz_{t-(l-m)} - \bzbar)^T \, .
\end{equation*}
Let $\tilgamma_{l-m} = (\tilGamma_{l-m})_{ij}$ and $\gamma_{l-m} = (\Gamma_{l-m})_{ij}$. Then, under the assumptions of \cref{thm:flat_loss}, it is shown in \citet{quinn1980order} that $n^{1/2 - \delta} (\tilgamma_{l-m} - \gamma_{l-m})$ converges almost surely to 0 for all $\delta > 0$. Thus, $n^{1/2 - \delta} (\tilgamma_{l-m} - \gamma_{l-m})$  also converges in probability to 0, which implies that $|\tilgamma_{l-m} - \gamma_{l-m}| = o_P(n^{-1/2 + \delta})$. These rates are not immediately applicable to $\hgamma_{l-m,p}$ so we next establish that $|\hgamma_{l-m,p} - \gamma_{l-m}| = o_P(n^{-1/2 + \delta})$ by simplifying $ \hat{\Gamma}_{l-m,p}$. We start by re-indexing the sum in $\hat{\Gamma}_{l-m,p}$ with $j = t-m$. Then,
\begin{equation*}
        \hat{\Gamma}_{l-m,p} = \frac{1}{n-p} \sum_{j = p-m+1}^{n-m} (\bz_{j} - \bzbar)(\bz_{j-(l-m)} - \bzbar)^T \, .   
\end{equation*}
With this formulation, it is easy to see that the difference between $\tilGamma_{l-m}$ and $\hGamma_{l-m,p}$ is that $\tilGamma_{l-m}$ has the additional indices $t = \{(l-m)+1, \dots, p-m \}$ and $t = \{n-m+1, \dots, n\}$ in the sum and also uses a scaling of $1/n$ instead of $1/(n-p)$. Note that the number of the extra terms does not grow with $n$ and instead only depends on $l,m,p$. Thus, these terms are $o(1/n)$. The explicit relationship between the two is given by 
\begin{equation}
    \label{eqn:autocov_equivalence}
    \tilGamma_{l-m} = \frac{n-p}{n} \left( \hGamma_{l-m,p} + \frac{1}{n-p} \left[ \sum_{j = (l-m)+1}^{p-m} (\bz_{j} - \bzbar)(\bz_{j-(l-m)} - \bzbar)^T + \sum_{j=n-m+1}^{n} (\bz_{j} - \bzbar)(\bz_{j-(l-m)} - \bzbar)^T \right] \right) \, .
\end{equation}
Using \cref{eqn:autocov_equivalence}, we can write
\begin{equation*}
    \begin{split}
        |\hgamma_{l-m,p} - \gamma_{l-m}| & = \left| \frac{n}{n-p} \tilgamma_{l-m} - o(n^{-1}) - \gamma_{l-m} \right| \\
        & = \left| \frac{n-p}{n-p} \tilgamma_{l-m} + \frac{p}{n-p} \tilgamma_{l-m} - \gamma_{l-m} - o(n^{-1}) \right| \\
        & \leq | \tilgamma_{l-m} - \gamma_{l-m} | + o_P(n^{-1}) + o(n^{-1}) \\
        & = o_P(n^{-1/2 + \delta}) \, ,
    \end{split}
\end{equation*}
where in the second to last line we used the fact that $\tilgamma_{l-m}$ converges in probability so $\tilgamma_{l-m} p/(n-p) = o_P(n^{-1})$. Since \cref{asm:invertibility} holds by the assumptions of \cref{thm:flat_loss} and the data is drawn from a continuous distribution, we can apply \cref{lem:loss_lipschitz} to see that
\begin{equation*}
\begin{split}
|\hLVAR(p) - \LVAR(p)| &\leq L \left\| \hat{G} - G^* \right\|_1 \\
|\hLVAR(p) - \LVAR(p)| &\leq L \frac{k^2(p+2)(p+1)}{2} o_P(n^{-1/2 + \delta}) \\
|\hLVAR(p) - \LVAR(p)| &= o_P(n^{-1/2 + \delta}) \, .
\end{split}
\end{equation*}
Note that the $\hat{G}$ is the input formed by the relevant autocovariance matrices used in the least squares loss and $G^*$ is the input consisting of the true autocovariance matrices as given in \cref{eqn:true_autocov_inputs}.
\end{proof}

\begin{proof}[Proof of \cref{lem:loss_lipschitz}]
For ease of notation, let 
\begin{align*}
    V(G) &:= \begin{bmatrix} \Gamma_1^0 & \dots & \Gamma_p^0 \end{bmatrix} \, , \\
    T(G) & := \begin{bmatrix} \Gamma_0^1 & \dots & \Gamma_{p-1}^1 \\ 
                                    \vdots & & \vdots \\
                                    (\Gamma_{p-1}^1)^{T} & \dots & \Gamma_0^{p} 
    \end{bmatrix} \, .
\end{align*} With this, we have that, \begin{equation*}
f(G) = \Tr\left( \Gamma_0^0 - V(G) T(G) V(G)^T \,. \right) \, .
\end{equation*}
To simplify notation further, we will suppress the dependence on G in the notation of $V(G), T(G)$ and use the shorthands $V := V(G)$ and $T := T(G)$. Let $G_i$ be the $i^{th}$ element of $G$. Using $ \deriv{\Tr(h(G))}{G_i} =  \Tr\left( \deriv{h(G)}{G_i} \right)$, we have that
\begin{equation*}
    \deriv{f(G)}{G_i}  = \Tr \left( \deriv{\Gamma_0^0}{G_i} - \deriv{V}{G_i} T^{-1} V^T - V \deriv{T^{-1}}{G_i} V^T - V T^{-1} \deriv{V^T}{G_i}
    \right) \, .
\end{equation*}
From~(59) of \cite{petersen2008matrix}, $\deriv{T^{-1}}{G_i} = - T^{-1} \deriv{T}{G_i} T^{-1}$. Thus, 
\begin{equation*}
    \deriv{f(G)}{G_i} = \Tr \left( \deriv{\Gamma_0^0}{G_i} -  \deriv{V}{G_i} T^{-1}  V^T + V T^{-1} \deriv{T}{G_i} T^{-1} V^T - V T^{-1} \deriv{V^T}{G_i}
    \right) \, .
\end{equation*}
Note that each of $\deriv{\Gamma_0^0}{G_i}, \deriv{V}{G_i}, \deriv{T}{G_i}$ are matrices containing $1$ in the entries of $\Gamma_0^0, V, T$ where $G_i$ is present and $0$ otherwise. 

While matrix multiplication and the trace are continuous functions in general, matrix inversion is a continuous function only over the set of invertible matrices. However since we have assumed that $G, \hat{G}$ are such that the inverses exist, we conclude that $f(G)$ is continuously differentiable.  By the mean value theorem and H\"older's inequality
\[
\left| f(\hat{G}) - f(G) \right| \leq \left\|\nabla f((1-\nu)G + \nu \hat{G})\right\|_{\infty} \left\|\hat{G} - G \right\|_{1} \, ,
\]
for some $\nu \in (0,1)$. Since $\deriv{f(G)}{G_i}$ is continuous for each $G_i$, it is bounded on any closed interval including between $G$ and $\hat{G}$. Thus $\left\|\nabla f((1-\nu)G + \nu \hat{G})\right\|_{\infty}$ is bounded and
\begin{equation*}
    \left| f(\hat{G}) - f(G) \right| \leq L \left\|\hat{G} -G \right\|_{1} \, ,
\end{equation*}
for some $L < \infty$.
\end{proof}

\begin{remark}
    As shown in the proof of \cref{thm:consistency}, the least squares estimator, $\hat{A}_p$, uses multiple estimates of the autocovariances $\Gamma_i$. Thus, it is necessary to define $G$ and $f(G)$ in \cref{lem:loss_lipschitz} to allow for multiple estimators of autocovariances $\Gamma_i$. In practice, however, the estimates of each of the autocvoariances $\Gamma_i$ used in $\hat{A}_p$ are nearly identical.
\end{remark}
\begin{remark}
        It is worth noting that if we define
    \begin{equation}
    \label{eqn:true_autocov_inputs}
        G^* = \begin{bmatrix} \vec{\Gamma}_{0} & \vec{\Gamma}_1 & \dots & \vec{\Gamma}_p & \vec{\Gamma}_{0} & \dots & \vec{\Gamma}_{0} & \vec{\Gamma}_{1} & \dots & \vec{\Gamma}_{1} & \vec{\Gamma}_{2} & \dots & \vec{\Gamma}_{2} & \dots & \vec{\Gamma}_{p-1}
        \end{bmatrix} \, ,
    \end{equation} where $\Gamma_i$ are the $i^{th}$ population autocovariances, then under \cref{asm:invertibility}, $f(G^*) = \LVAR(p)$. That is, $f(G^*)$ is the expected squared error loss at fitted order $p$.
\end{remark}
\begin{remark}
    We require $G$ and $\hat{G}$ to be such that the inverses in $f(G), f(\hat{G}), f((1-\nu)G + \nu \hat{G})$ exist for all $\nu \in (0,1)$, as the inverse function is only continuous over the set of invertible matrices. In our theoretical analysis of \cref{thm:consistency}, this holds by \cref{asm:invertibility} and the fact that $\hat{G}$ is generated from a continuous distribution so that the inverse in $f(\hat{G})$ exists with probability 1 and similarly the inverse in $ f((1-\nu)G + \nu \hat{G})$ is also exists with probability 1 for all $\nu \in (0,1)$.
\end{remark}

To better understand teh structure of $\deriv{\Gamma_0^0}{G_i}, \deriv{V}{G_i}, \deriv{T}{G_i}$ mentioned in the proof of \cref{lem:loss_lipschitz}, we consider two examples. 
\begin{example}
   For the first example consider taking the partial derivative with respect to the first component, $G_1 = (\Gamma_0^0)_{1,1}$. That is, $G_1$ is the $(1,1)$ entry of the $\Gamma_0^0$ parameter. Then, the partial derivatives are
\begin{align*}
    \deriv{\Gamma_0^0}{G_1} & = \begin{bmatrix} 1 & 0 & \dots & 0 \\ 
                                        0 & 0 & \dots & 0 \\ 
                                        \vdots & \vdots & & \vdots \\ 
                                        0 & 0 & \dots & 0
                                \end{bmatrix} \, , \\
        \deriv{V}{G_1} & = 0 \, , \\
        \deriv{T}{G_1} & = 0 \, .
\end{align*}
\end{example}
\begin{example}
    For the second example consider taking the partial derivative with respect to the $(k^2(2p+1) + 1)$-th component in $G$. We have that $G_{k^2(2p+1) + 1} = (\Gamma_{1}^{1})_{1,1}$. The resulting partial derivatives are
\begin{align*}
    \deriv{\Gamma_0^0}{G_{k^2(2p+1) + 1}} & = 0 \, , \\
        \deriv{V}{G_{k^2(2p+1) + 1}} & = 0 \, , \\
        \deriv{T}{G_{k^2(2p+1) + 1}} & = \begin{bmatrix}
          0_{k \times k} & \left[\begin{array}{cccc}
         1 & 0 & \dots & 0 \\
         \vdots & & \dots & \\
         0 & 0 & \dots & 0 \\
  \end{array}\right] & \dots & 0_{k \times k} \\
         \left[\begin{array}{cccc}
         1 & 0 & \dots & 0 \\
         \vdots & & \dots & \\
         0 & 0 & \dots & 0 \\
  \end{array}\right] &  0_{k \times k} & \dots & 0_{k \times k} \\
         \vdots & & \ddots & \vdots \\
         0_{k \times k} & 0_{k \times k} & \dots & 0_{k \times k} 
         \end{bmatrix}
         \, .
\end{align*}
\end{example}

\section{Additional simulation results}\label{app:more_sims}
\begin{figure}[ht!]
\centering
\includegraphics[width=0.7\linewidth]{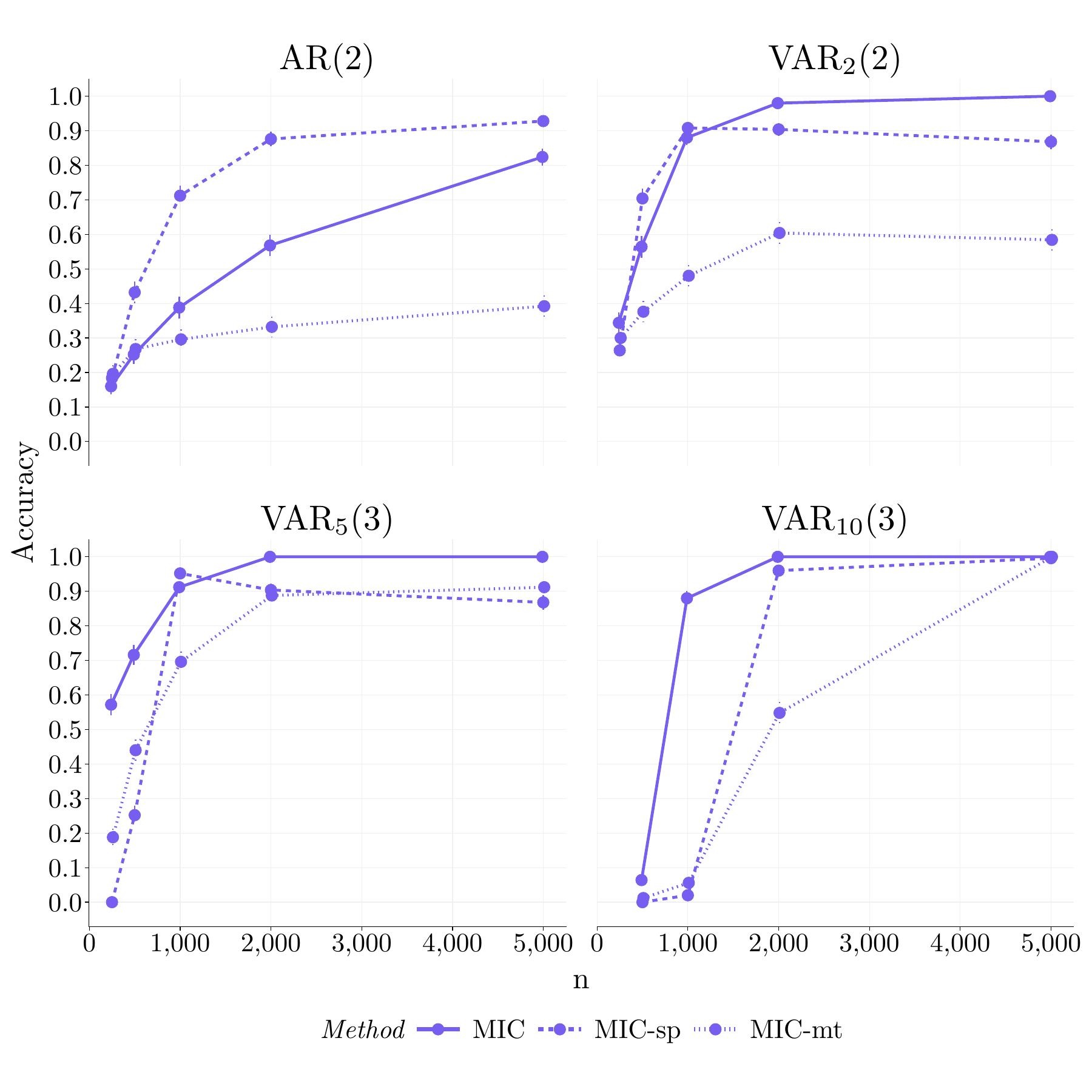}
\caption[Diagonal Gaussian errors MIC comparison]{\textbf{Diagonal Gaussian errors}. Simulation results comparing accuracy of different MIC order selection methods. Vertical lines indicate standard errors. }
\label{fig:MICcomp_diagGaus_sims}
\end{figure}

\begin{figure}[ht!]
    \centering
    \includegraphics[width=0.4\linewidth]{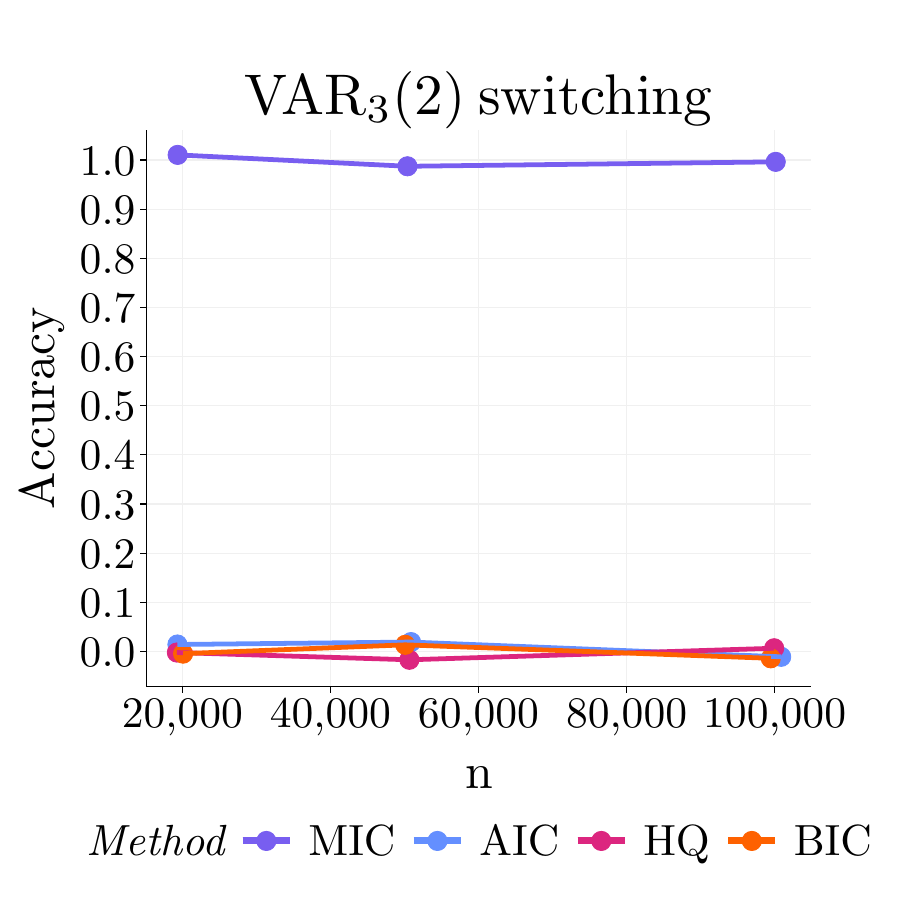}
\caption[$\VAR_3(2)$ switching large sample simulation results]{\textbf{VAR}$\mathbf{_3(2)}$\textbf{ switching}. Large sample size simulation results. Vertical lines indicate standard errors. Points have been jittered by 0.02 in the vertical and 1000 in the horizontal directions to improve readability.}
    \label{fig:var3_2_switch_large_sims}
\end{figure}

\begin{figure}[ht!]
\centering
\includegraphics[width=\linewidth]{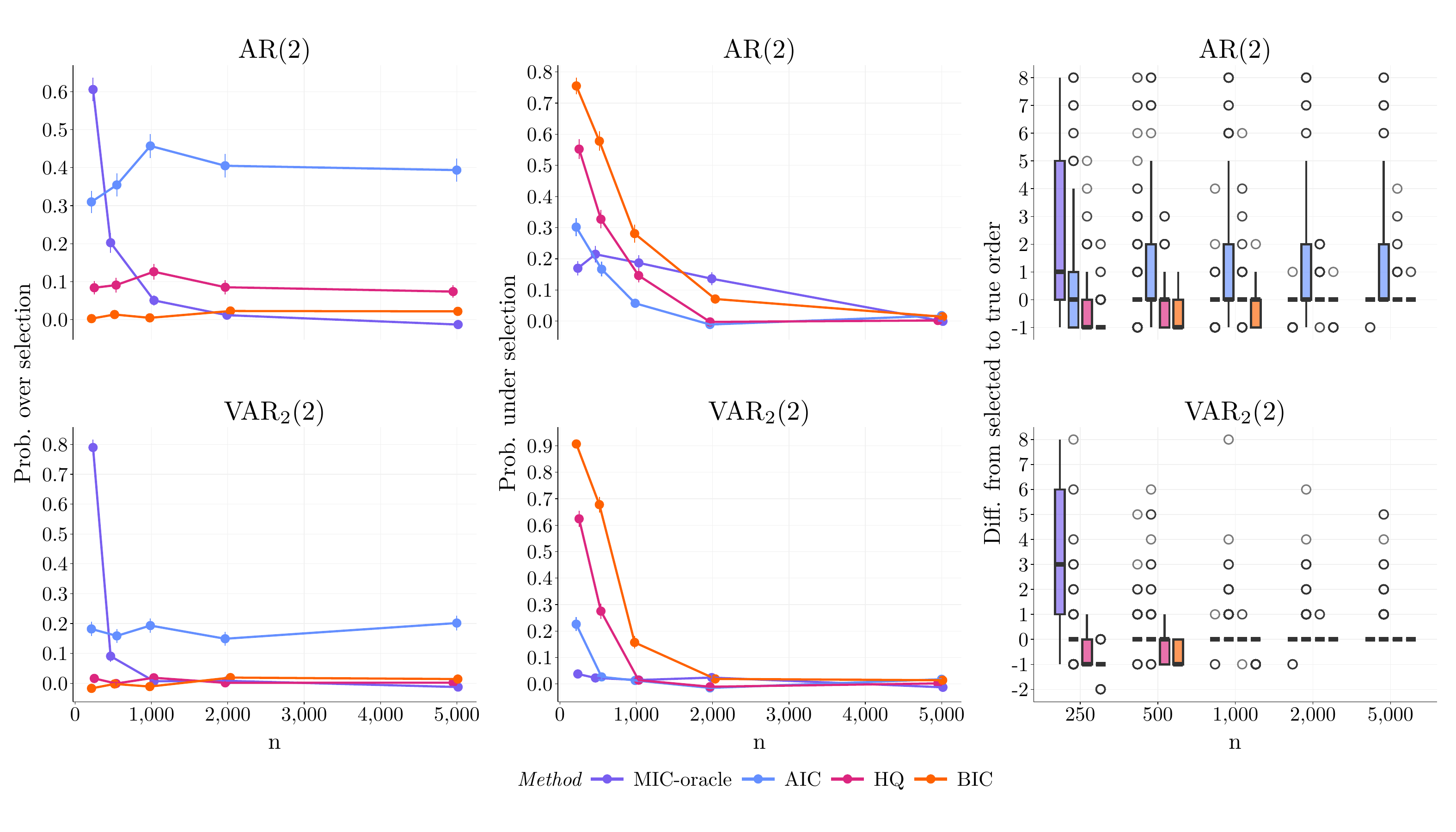}
\caption[Diagonal Gaussian errors over / under selection, small k]{\textbf{Diagonal Gaussian errors}. Simulation results for over and under selection of order with diagonal Gaussian errors. Vertical lines indicate standard errors. Points in the over and under selection plots have been jittered by 0.02 in the vertical and 50 in the horizontal directions to improve readability.}
\label{fig:over_under_diagGaus_sims1}
\end{figure}

\begin{figure}[ht!]
\centering
\includegraphics[width=\linewidth]{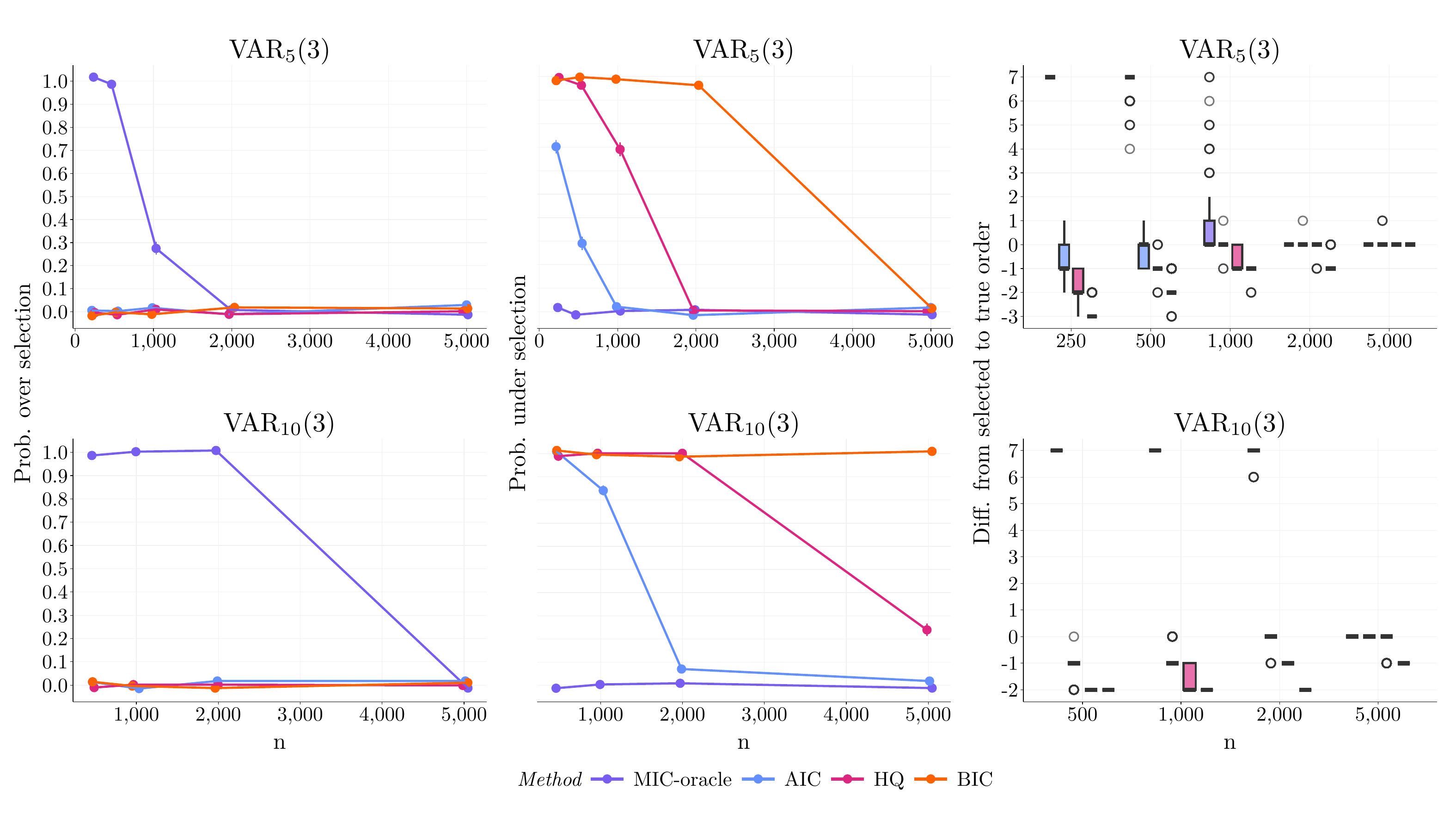}
\caption[Diagonal Gaussian errors over / under selection, large k]{\textbf{Diagonal Gaussian errors}. Simulation results for over and under selection of order with diagonal Gaussian errors. Vertical lines indicate standard errors. Points in the over and under selection plots have been jittered by 0.02 in the vertical and 50 in the horizontal directions to improve readability.}
\label{fig:over_under_diagGaus_sims2}
\end{figure}

\begin{figure}[ht!]
\centering
\includegraphics[width=\linewidth]{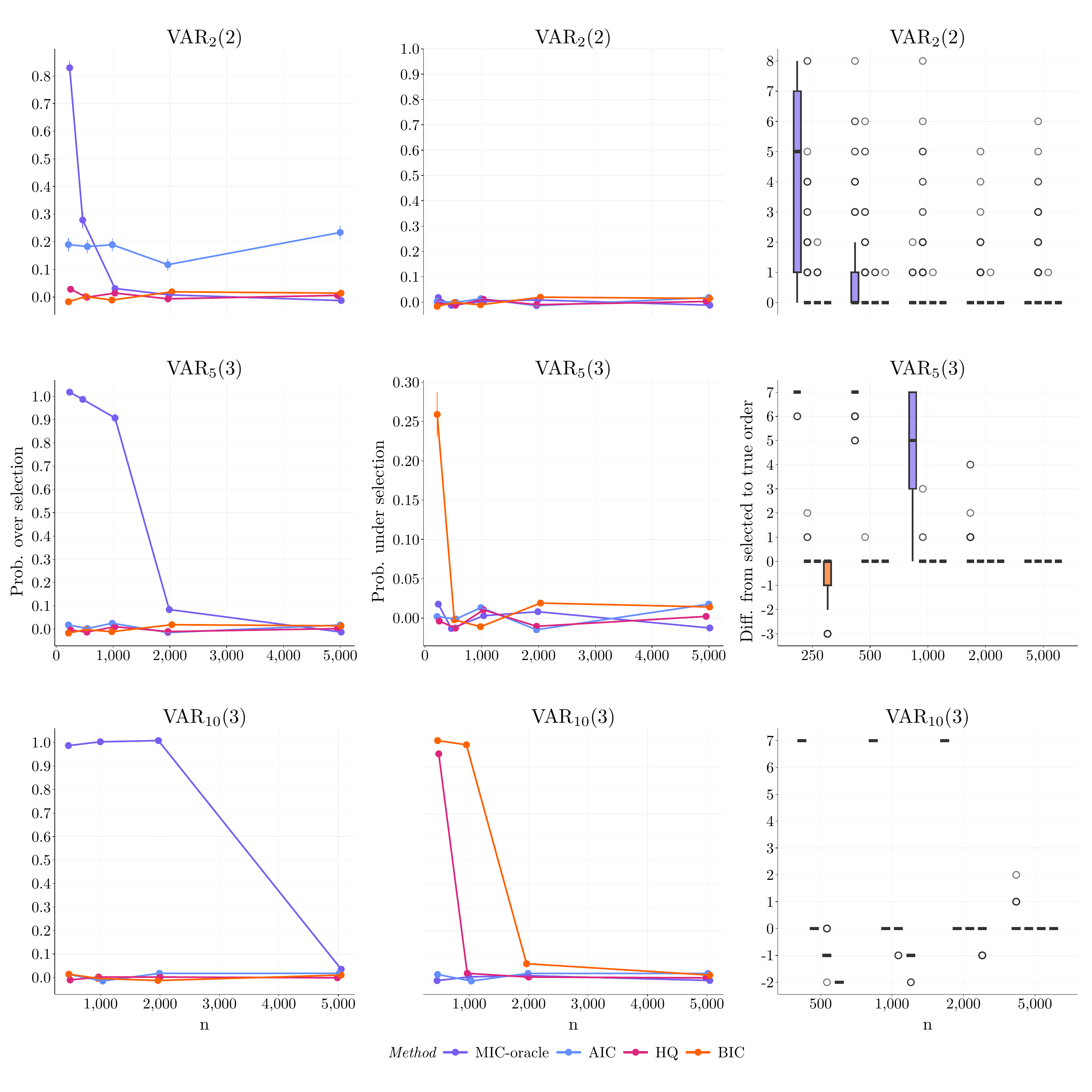}
\caption[Non-diagonal Gaussian errors over / under selection]{\textbf{Non-diagonal Gaussian errors}. Simulation results for over and under selection of order with non-diagonal Gaussian errors. Vertical lines indicate standard errors. Points in the over and under selection plots have been jittered by 0.02 in the vertical and 50 in the horizontal directions to improve readability.}
\label{fig:over_under_nonDiagGaus_sims}
\end{figure}

In this section, we present additional simulation results. In \cref{fig:MICcomp_diagGaus_sims} we compare order selection by minimizing MIC with $\lambda_{\mathrm{ST}}$ (denoted as MIC) to the alternative procedures, MIC-sp and MIC-mt, from \cref{sec:alt_lambdas}. Recall that MIC-sp uses MIC with a penalty $\lambda_{\mathrm{sp}}$ that is chosen by using a 70-30 train-test split of the data while MIC-mt also uses a 70-30 split but chooses the order $0, \dots, p_{\mathrm{max}}$ that minimizes the test error. Overall we see that MIC offers the best performance over all settings and sample sizes. In \cref{fig:var3_2_switch_large_sims} we study the performance of each order selection method for large sample sizes in the $\VAR_{3}(2)$ switching simulation setting. MIC is the only method that consistently estimates the true order for all sample sizes. Lastly, as discussed in \cref{sec:over_under_selection}, we compare the likelihood of over and under selection of the true model order for MIC using an oracle $\lambda$ to each of the competing order selection methods. The results in \cref{fig:over_under_diagGaus_sims1,fig:over_under_diagGaus_sims2,fig:over_under_nonDiagGaus_sims} show that MIC-oracle tends to over select the true order while the alternative methods tend to under select the true order. Moreover, as seen in \cref{fig:over_under_nonDiagGaus_sims}, the probability of over selection in MIC-oracle may be worse than the corresponding probability of under selection from AIC, BIC, and HQ when the dimension is large and the errors are non-diagonal Gaussian.

\clearpage

\section{Daily realized stock variances}\label{app:finance_application}
\cref{tbl:stocks_descr} shows the stocks and their corresponding descriptions that were included in the financial forecasting analysis from \cref{sec:stock_analysis}. A check mark in the $k = 16$ or $k = 7$ column indicates that the stock was included in the corresponding analysis shown in \cref{tbl:wmsfe_finance}.

\begin{table}[!hb]
\begin{center}
\begin{tabular}{ l l c c}
\toprule
 Stock & Description & $k = 16$ & $k = 7$ \\ 
 \midrule
AEX & Amsterdam Exchange Index & \checkmark & \\
AORD & All Ordinaries Index & \checkmark & \\
BFX & Belgium Bell 20 Index & \checkmark & \\
BVSP & BOVESPA Index & \checkmark & \\
DJI & Dow Jones Industrial Average & \checkmark & \\
FCHI & Cotation Assist´ee en Continu Index & \checkmark & \checkmark \\
FTSE & Financial Times Stock Exchange Index 100 & \checkmark & \checkmark \\
GDAXI & Deutscher Aktienindex & \checkmark & \checkmark \\
HSI & HANG SENG Index & \checkmark & \checkmark \\
IXIC & Nasdaq stock index & \checkmark & \\
KS11 & Korea Composite Stock Price Index & \checkmark & \checkmark \\
MXX & IPC Mexico & \checkmark & \\
N225 & Tokyo stock exchange index & &  \checkmark \\
RUT & Russel 2000 & \checkmark & \\
SPX & Standard \& Poor’s 500 market index & \checkmark & \checkmark \\
SSMI & Swiss market index & \checkmark & \\
STOXX50E & EURO STOXX 50 & \checkmark & \\
 \bottomrule
\end{tabular}
\end{center}
    \caption{Stocks analyzed in financial application}
    \label{tbl:stocks_descr}
\end{table}

\end{document}